\documentclass[superscriptaddress,groupedaddress,nofootnoteinbib,12pt]{article}
\usepackage{color}
\usepackage{graphicx}
\usepackage{dcolumn}
\usepackage{bm}
\usepackage{amssymb}
\usepackage{amsmath}
\usepackage{sectsty}
\usepackage{colortbl}
\usepackage{latexsym}
\usepackage{float}
\usepackage{ifthen}
\usepackage{caption,subfig}
\usepackage{enumerate}
\usepackage{url}
\usepackage{caption,subfig}	
\usepackage{jcappub}

\newcommand{\eff}{{\rm eff}}

\def\be{\begin{equation}}
\def\ee{\end{equation}}
\def\ba{\begin{eqnarray}}
\def\ea{\end{eqnarray}}
\def\beq{\begin{eqnarray}}
\def\eeq{\end{eqnarray}}

\def\mpl{M_{\rm Pl}}

\def\d{\mathrm{d}}

\def\L*{{\cal L}_*}
\def\L{\mathcal{L}}
\def\({\left(}
\def\){\right)}

\def\<{\langle}
\def\>{\rangle}

\def\cs2{c_{s}^{2}}

\def\be{\begin{equation}}
\def\ee{\end{equation}}
\def\ba{\begin{eqnarray}}
\def\ea{\end{eqnarray}}
\def\beq{\begin{eqnarray}}
\def\eeq{\end{eqnarray}}

\def\mpl{M_{\rm Pl}}

\def\d{\mathrm{d}}

\def\L*{{\cal L}_*}
\def\L{\mathcal{L}}
\def\({\left(}
\def\){\right)}

\def\<{\langle}
\def\>{\rangle}

 \def\be   {\begin{equation}}   \def\ee   {\end{equation}}

 \def\ba  {\begin{eqnarray}}   \def\ea  {\end{eqnarray}}

\renewcommand{\thesubsection}{\arabic{section}.\arabic{subsection}}

\begin{document}
\hspace{5.2in} \mbox{YITP-15-78, IPMU15-0162}\\\vspace{-1.03cm} 
\title{Matter coupling in partially constrained vielbein formulation of massive gravity}

\author{Antonio De Felice$^{a}$, A. Emir G\"umr\"uk\c c\"uo\u glu$^{b}$, Lavinia Heisenberg$^{c}$, Shinji Mukohyama$^{a,d}$}
\affiliation{$^{a}$Yukawa Institute for Theoretical Physics, Kyoto University, \\
Kyoto 606-8502, Japan}
\affiliation{$^{b}$School of Mathematical Sciences, University of Nottingham, \\
University Park, Nottingham, NG7 2RD, UK}
\affiliation{$^{c}$Institute for Theoretical Studies, ETH Zurich\\
Clausiusstrasse 47, 8092 Zurich, Switzerland}
\affiliation{$^{d}$Kavli Institute for the Physics and Mathematics of the Universe,\\
Todai Institutes for Advanced Study, University of Tokyo (WPI),\\
5-1-5 Kashiwanoha, Kashiwa, Chiba 277-8583, Japan}

\emailAdd{antonio.defelice@yukawa.kyoto-u.ac.jp}
\emailAdd{emir.gumrukcuoglu@nottingham.ac.uk} 
\emailAdd{lavinia.heisenberg@eth-its.ethz.ch}
\emailAdd{shinji.mukohyama@ipmu.jp}

\abstract{We consider a consistent linear effective vielbein matter coupling without introducing the Boulware-Deser ghost in ghost-free
massive gravity. This is achieved in the partially constrained vielbein formulation. We first introduce the formalism and prove the absence of ghost at all scales. As next we investigate the cosmological
application of this coupling in this new formulation. We show that even if the background evolution accords with the metric formulation, the perturbations display important different
features in the partially constrained vielbein formulation. We study the cosmological perturbations of the two branches of solutions separately. The tensor perturbations coincide with those in the metric formulation. Concerning the vector and scalar perturbations, the requirement of absence of ghost and gradient instabilities yields slightly different allowed parameter space. }

\maketitle

\section{Introduction}

The existence of a graviton mass is an unavoidable fundamental question from a theoretical perspective. The pioneering work of Fierz and Pauli marked an important progress \cite{Fierz:1939ix}. It is the unique mass term at the linear level, which does not lead to the presence of ghosts at the classical level in the theory. Even though theoretically completely consistent, this theory of massive gravity suffers unfortunately from the vDVZ discontinuity \cite{vanDam:1970vg,Zakharov:1970cc}, i.e., General Relativity predictions are not recovered in the limit of massless gravitons. The main reason for this is the existence of an additional scalar degree of freedom in massive gravity that couples to the trace of the energy momentum tensor. Obviously the vDVZ discontinuity is just an artifact of the linear approximation. The effects of massive gravity might be cloaked by the Vainshtein mechanism, where the helicity-0 mode interactions become appreciable to freeze out the field fluctuations on small scales. Nevertheless, such non-linear extensions usually have the Boulware-Deser ghost instability when non-trivial backgrounds are considered. Recently, a ghost-free non-linear theory of massive gravity was pushed forward \cite{deRham:2010ik,deRham:2010kj,Hassan:2011vm,Hassan:2011hr}, which has sparked a renewed interest in massive gravity, specially for its possible application in cosmology. \\

The phenomenological exploration of the theory has triggered a knock-on effect for extensions of the standard formulation. First of all a theoretically well expected no-go result was faced soon \cite{PhysRevD.84.124046}. The same condition to remove the Boulware-Deser ghost enforces the absence of flat FLRW solutions in the case of flat fiducial metric. 
Although a self-accelerating open FLRW solution exists \cite{Gumrukcuoglu:2011ew}, it suffers from non-linear ghost instability \cite{PhysRevLett.109.171101, DeFelice:2013awa}.
Trying to bypass this difficulty has initiated considerations of more general fiducial metrics, which sadly was soon doomed to have Higuchi type instabilities in the case of dS reference metric \cite{Fasiello:2012rw,Langlois:2012hk} and to protect against acceleration 
in the case of AdS reference metric \cite{MartinMoruno:2013gq}. Moreover, the self-accelerating branch solutions are not affected by the choice of the reference metric and they all behave like the open solutions above \cite{Gumrukcuoglu:2011zh, Gumrukcuoglu:2012wt}.
This on the other hand has launched extensions of the theory by adding new degrees of freedom \cite{Huang:2012pe,PhysRevD.87.064037,DeFelice:2013tsa}. The study of quantum properties of the potential interactions has motivated a new branch of research for consistent matter couplings in the theory. The requirement of maintaining the special structure of the potential interactions intact at the quantum level yields severe restrictions on the possible couplings. The phenomenological consequences of massive (bi-)gravity depend highly non-trivially on the way how matter fields couple to the massive graviton  \cite{Khosravi:2011zi, Akrami:2012vf,Akrami:2013ffa,Tamanini:2013xia,Akrami:2014lja, deRham:2014naa,deRham:2014fha,Yamashita:2014fga,Noller:2014sta,Schmidt-May:2014xla,Enander:2014xga,Solomon:2014iwa,Soloviev:2014eea,Heisenberg:2014rka,Huang:2015yga,Heisenberg:2015wja,Blanchet:2015sra,Blanchet:2015bia,Bernard:2015gwa,Mukohyama:2014rca,Lagos:2015sya}. The above mentioned difficulties in the cosmological setup arise only in the case of the minimal coupling to one of the metrics, which can be avoided by considering a matter coupling through a very specific admixture of the dynamical and fiducial metric \cite{deRham:2014naa}. Considerations of this direction usually reintroduce the Boulware-Deser ghost. The important question there is at what scale this ghostly degree of freedom enters. In the context of the special composite effective metric proposed in \cite{deRham:2014naa}, the first non-trivial non-vanishing leading order vector-scalar-matter interactions typically reintroduce the Boulware-Deser ghost assuming full local Lorentz symmetry \cite{Matas:2015qxa}. However, this does not preclude us from considering these interactions as a consistent effective field theory with the cut-off given by the mass of the ghost. Not only avoids this specific coupling the no-go result for the flat FLRW solutions, it does it so in a very specific way. The mass of the Boulware-Deser ghost on the FLRW background is infinite. This is because the ghostly vector-scalar-matter interactions do not contribute to the background due to the symmetry of FLRW. \\

Of course it would be more tempting to construct matter couplings in which the absence of the Boulware-Deser ghost is realised fully non-linearly. Not only would this enable the viability of the Vainshtein mechanism, but also enlarge the scale of applicability. A possible way out was argued in \cite{Hinterbichler:2015yaa} by switching to the unconstrained vielbein formulation. The argumentation is simple and elegant. In the case of non-derivative coupling the resulting Hamiltonian is linear in the effective lapse and shift. In the boosted ADM decomposition of the vielbein one can easily express the effective lapse and shift in terms of the lapses and shifts of the two vielbeins. After using the equations of motion for the shift one can integrate out the boost parameters resulting in effective lapse and shift that depends only linearly on the lapses and shifts of the two vielbeins. Hence, the linearity in the lapse and shift enforce the first class primary constraints that remove the Boulware-Deser ghost. So far so good, this argumentation has however an unfortunate loophole. The existence of the secondary constraints is taken for granted. Nevertheless, this turns out to be not the case. In \cite{deRham:2015cha}, it was shown explicitly that the Hamiltonian becomes highly non-linear in the lapses once the rotations are integrated out as well, which reflects the absence of the secondary constraints along the line of the analysis performed in \cite{Hinterbichler:2015yaa}. Thus, also in the unconstrained vielbein formulation the Boulware-Deser ghost remains persistent in the non-minimal coupling.\\

In this work we will consider yet another formulation of the coupling, which we call partially constrained vielbein formulation. For this purpose we shall go along the lines of a recent study in \cite{DeFelice:2015hla},  where the goal was to remove the unwanted, unstable degrees of freedom of dRGT theory, by only keeping its tensor modes. The advantage of this formulation is the absence of the Boulware-Deser ghost to all orders beyond the decoupling limit. Any general vielbein can be decomposed into a Lorentz boost and rotation of a triangular vielbein. In \cite{Hinterbichler:2015yaa}, the integration of the $(D-1)$ boost parameters in $D$ dimensions yields a linear Hamiltonian, however the integrations of the additional $(D-1)(D-2)/2$ rotation parameters yields a Hamiltonian highly non-linear in the lapses \cite{deRham:2015cha}. Our partially constrained vielbeins are constructed such that the rotation parameters do not reintroduce the non-linearities in the lapses. Hence, the argumentation of \cite{Hinterbichler:2015yaa} applies exactly and the loophole found in \cite{deRham:2015cha} is removed. We shall further study the cosmological consequences of the partially constrained formulation of the matter coupling. For this we will assume the same background evolution as for the metric perturbations with vanishing boost parameters. However, important differences arise at the level of the perturbations. We shall first present the dRGT theory in the partially constrained formulation in Section \ref{sec:dRGT} and state our convenient conventions and notations.
We then prove the absence of the Boulware-Deser ghost related to the non-minimal coupling in this new formulation in Section \ref{absence_ghost}. As next we study the background equations of motion on FLRW space-time in Section \ref{sec:background_evolution} with the presence of two branches of solutions. Similarly as in the previous studies of the effective coupling the new coupling enables us to avoid the no-go theorem for flat FLRW solutions. Finally, we shall investigate in detail the stability of the perturbations in the two branches in Section \ref{sec:perturbations1}.


\section{Partially constrained vielbein formulation}
\label{sec:dRGT}

The original formulation of de Rham-Gabadadze-Tolley (dRGT) massive gravity in the metric language is mathematically cumbersome due to the presence of matrix square roots. 
Not only is the Hamiltonian analysis hindered but also the cosmological perturbations. For instance to prove the existence of the primary constraint one has to perform a highly non-trivial field redefinition of the shift. Since the vielbein is like the square root of the metric, the study of the potential interactions simplify drastically when one uses the vielbein formulation. We can express the two metrics by the corresponding vielbeins as 
\begin{equation}
 g_{\mu\nu}  =  \eta_{\cal AB} e^{\cal A}{}_{\mu} e^{\cal B}{}_{\nu}  \qquad \text{and} \qquad
 f_{\mu\nu}  =  \eta_{\cal AB} E^{\cal A}{}_{\mu} E^{\cal B}{}_{\nu}\,.
\label{eq:metrics}
\end{equation}
Hereafter, indices $\mathcal{A},\mathcal{B},\mathcal{C},\mathcal{D}$ and $\mu,\nu,\rho,\sigma$ run from $0$ to $3$, but indices $I,J,K,L$ and $i,j,k,l$ run from $1$ to $3$. In terms of this Einstein-Cartan formulation, the potential interactions simply correspond to all possible wedge products of the vielbein of the dynamical metric with the vielbein of the second metric
\begin{eqnarray}
 {\cal L}_0 & = & \frac{1}{24}
  \epsilon^{\mu\nu\rho\sigma}\epsilon_{\cal ABCD}
  E^{\cal A}{}_{\mu}E^{\cal B}{}_{\nu}E^{\cal C}{}_{\rho}E^{\cal D}{}_{\sigma}\,,
  \nonumber\\
 {\cal L}_1 & = & \frac{1}{6} 
  \epsilon^{\mu\nu\rho\sigma}\epsilon_{\cal ABCD}
  E^{\cal A}{}_{\mu}E^{\cal B}{}_{\nu}E^{\cal C}{}_{\rho}e^{\cal D}{}_{\sigma}\,,
  \nonumber\\
 {\cal L}_2 & = & \frac{1}{4}
  \epsilon^{\mu\nu\rho\sigma}\epsilon_{\cal ABCD}
  E^{\cal A}{}_{\mu}E^{\cal B}{}_{\nu}e^{\cal C}{}_{\rho}e^{\cal D}{}_{\sigma}\,,
  \nonumber\\
 {\cal L}_3 & = & \frac{1}{6}
  \epsilon^{\mu\nu\rho\sigma}\epsilon_{\cal ABCD}
  E^{\cal A}{}_{\mu}e^{\cal B}{}_{\nu}e^{\cal C}{}_{\rho}e^{\cal D}{}_{\sigma}\,,
  \nonumber\\
 {\cal L}_4 & = & \frac{1}{24}
  \epsilon^{\mu\nu\rho\sigma}\epsilon_{\cal ABCD}
  e^{\cal A}{}_{\mu}e^{\cal B}{}_{\nu}e^{\cal C}{}_{\rho}e^{\cal D}{}_{\sigma}\,,
  \label{eqn:def-gravitonmassterm}
 \end{eqnarray}
with the Levi-Civita symbol normalized as $\epsilon_{0123}=-\epsilon^{0123}=1$. The Einstein-Hilbert term in pure General Relativity written in the vielbein formulation is trivial in the sense, that it corresponds to introducing new non-dynamical fields together with gauge invariances, exactly in the same philosophy as the St\"uckelberg trick. It admits four diffeomorphisms and six local Lorentz transformations. Introducing the potential interactions breaks these symmetries. In bigravity, the two copies of local Lorentz transformations and diffeomorphisms are broken down to a single copy. Of course one can introduce St\"uckelberg fields in order to restore the broken Lorentz and diffeomorphism invariance.
The vielbein formulation of the bi-gravity theory is dynamically equivalent to the metric formulation if one imposes the symmetric vielbein condition, which is a direct consequence of the equations of motion for the Lorentz St\"uckelberg field.

In this work we will be concentrating on massive gravity, i.e. the case with non-dynamical fiducial vielbein $E^{\cal A}{}_{\mu}$ together with its dual basis $E_{\cal A}{}^{\mu}$
satisfying 
\begin{equation}
 E^{\cal A}{}_{\mu}E_{\cal A}{}^{\nu} = \delta_{\mu}^{\nu}, \quad E^{\cal A}{}_{\mu}E_{\cal B}{}^{\mu} = \delta^{\cal A}_{\cal B}\,.
\end{equation}
We also introduce the dual basis $e_{\mathcal{A}}^{\ \mu}$ for the dynamical physical vielbein $e^{\mathcal{A}}_{\ \mu}$ as
\begin{equation}
 e^{\cal A}{}_{\mu}e_{\cal A}{}^{\nu} = \delta_{\mu}^{\nu}\,, \quad 
  e^{\cal A}{}_{\mu}e_{\cal B}{}^{\mu} = \delta^{\cal A}_{\cal B}\,.
\end{equation}
As already stated, the graviton mass term as well as the Einstein-Hilbert action are invariant under the overall local Lorentz transformation of the two vielbeins. We can fix the gauge freedom associated with the boost part of the overall local Lorentz transformation by demanding that the fiducial vielbein is of the so called Arnowitt-Deser-Misner (ADM) form:
\begin{equation}
 E^{\mathcal{A}}_{\ \mu} = 
  \left(
   \begin{array}{cc}
    M & 0\\
    M^kE^I_{\ k} & E^I_{\ j}
    \end{array}
       \right). \label{eqn:ADMvielbein}
\end{equation}
Here, $M$, $M^i$ and $E^I_{\ j}$ are the fiducial lapse, the fiducial shift and the fiducial spatial vielbein since the corresponding $4$-dimensional fiducial metric $f_{\mu\nu}$ becomes
\begin{equation}
 f_{\mu\nu}dx^{\mu}dx^{\nu} = -M^2dt^2 + f_{ij}(dx^i+M^idt)(dx^j+M^jdt), \quad
  f_{ij} = \delta_{IJ}E^I_{\ i}E^J_{\ j}. 
\end{equation}
On the other hand, after setting the fiducial vielbein to be of the ADM form (\ref{eqn:ADMvielbein}), in general the physical vielbein cannot be cast into the ADM form simultaneously. Instead, by using the known fact that any vielbein can be written in the so called boosted ADM form, we parametrize $e^{\mathcal{A}}_{\ \mu}$ by the physical lapse $N$, the physical shift $N^i$, the physical spatial vielbein $e^I_{\ j}$ and the boost parameter $b_I$ as
\begin{equation}
 e^{\mathcal{A}}_{\ \mu} = 
  \left(
   \begin{array}{cc}
    N\gamma + N^kb_Le^L_{\ k} & b_Le^L_{\ j}\\
    Nb^I+N^ke^L_{\ k}\left(\delta^I_L+\frac{b^Ib_L}{\gamma+1}\right) & 
     e^L_{\ j}\left(\delta^I_L+\frac{b^Ib_L}{\gamma+1}\right)
    \end{array}
       \right), \label{eqn:boosted-ADMvielbein}
\end{equation}
where $b^I=\delta^{IJ}b_J$ and $\gamma=\sqrt{1+b^Ib_I}$. This is not a gauge choice nor a physical condition but simply a reprametrization of a general vielbein. In terms of these variables the physical metric is written as
\begin{equation}
 g_{\mu\nu}dx^{\mu}dx^{\nu} = -N^2dt^2 + g_{ij}(dx^i+N^idt)(dx^j+N^jdt), \quad
  g_{ij} = \delta_{IJ}e^I_{\ i}e^J_{\ j}. 
\end{equation}

After fixing the boost part of the overall local Lorentz transformation as (\ref{eqn:ADMvielbein}), we further impose the following symmetric condition on the $3$-dimensional spatial vielbein:
\begin{equation}
 Y_{IJ} = Y_{JI}, \label{eqn:YIJ=YJI}
\end{equation}
where $Y_{IJ} \equiv E_I^{\ l}\delta_{JK}e^K_{\ l}$. The condition (\ref{eqn:YIJ=YJI}) is not a gauge condition but is a physical condition since it is invariant under the spatial rotational part of the overall local Lorentz transformation. Imposing (\ref{eqn:YIJ=YJI}) thus modifies the original formulation and defines a different formulation. Since the physical condition (\ref{eqn:YIJ=YJI}) treats the spatial components and the temporal component of the vielbein in a different way, this formulation potentially violates local Lorentz invariance in the gravity sector. We call massive gravity with the symmetric condition (\ref{eqn:YIJ=YJI}) a {\it partially constrained vielbein formulation} of massive gravity.

We thus have (at least) four formulations of the dRGT massive gravity: the metric formulation, the constrained vielbein formulation, the unconstrained vielbein formulation and the partially constrained vielbein formulation. All of the four formulations of the dRGT theory are equivalent on shell, i.e. after imposing equations of motion. This is because in the unconstrained or partially constrained vielbein formulation, the equations of motion set the $4$-dimensional vielbein to be symmetric on shell. As we shall see in the following sections, this is no longer the case if we introduce composite coupling to matter fields. The advantage of the partially constrained vielbein formulation is that, unlike other formulations, matter fields can consistently couple to a composite vielbein without turning on a (would-be) Boulware-Deser ghost at all scales.

We have already fixed the boost part of the overall Lorentz transformation by demanding the ADM form as in (\ref{eqn:ADMvielbein}). On the other hand, we have not fixed the spatial rotational part of the overall Lorentz transformation. While it is not necessary, for practical purposes it is sometimes convenient to fix the spatial rotational part as well, e.g. by introducing a fixed non-singular $3\times 3$ matrix ${\cal E}^{Jk}$ and imposing the condition 
\begin{equation}
  E^I_{\ k}{\cal E}^{Jk} = E^J_{\ k}{\cal E}^{Ik}. 
\end{equation}
Because of the standard polar decomposition (applied to the $3\times 3$ matrix $E^I_{\ k}{\cal E}^{Jk}$), different choices of ${\cal E}^{Jk}$ simply correspond to different choices of the spatial rotational part of the overall local Lorentz transformation, on which the graviton mass term and the Einstein-Hilbert action (as well as the matter action, even with the composite coupling discussed below) do not depend. For example, for the analysis of perturbations around a particular background, it is probably most convenient to choose ${\cal E}^{Jk}$ to be the background value of $\delta^{JL}E_L^{\ k}$.

On introducing the variables $Y_{\cal B}{}^{\cal A}$ and $X_{\cal B}{}^{\cal A}$ as
\begin{equation}
 Y_{\cal B}{}^{\cal A} \equiv E_{\cal B}^{\ \mu}e^{\cal A}_{\ \mu}, 
  \quad
  X_{\cal B}{}^{\cal A} \equiv e_{\cal B}^{\ \mu}E^{\cal A}_{\ \mu}, 
\end{equation}
we can rewrite the potential interactions (\ref{eqn:def-gravitonmassterm}) simply as
\begin{eqnarray}
 {\cal L}_0 & = & - \sqrt{-f}\,, \nonumber\\
 {\cal L}_1 & = & - \sqrt{-f} Y_{\cal A}{}^{\cal A}\,, \nonumber\\
 {\cal L}_2 & = & - \frac{1}{2}\sqrt{-f}
  \left( Y_{\cal A}{}^{\cal A}Y_{\cal B}{}^{\cal B} - Y_{\cal A}{}^{\cal B}Y_{\cal B}{}^{\cal A}\right)\,, \nonumber\\
 {\cal L}_3 & = & - \sqrt{-g} X_{\cal A}{}^{\cal A}\,, \nonumber\\
 {\cal L}_4 & = & - \sqrt{-g}\,.
\end{eqnarray}
In the absence of the composite matter coupling introduced in (\ref{eq:tetradeff-def}) below, i.e. if matter fields minimally couple to the physical metric $g_{\mu\nu}$ then the symmetric vielbein ($Y^{\cal AB}=Y^{\cal BA}$) is a solution to the equations of motion (as in the unconstrained vielbein formulation) and thus the partially constrained vielbein formulation agrees with the symmetric vielbein formulation and the metric formulation.

Concerning the coupling to the matter fields, we will only allow non-derivative couplings through a specific composite effective metric proposed in \cite{deRham:2014naa, Noller:2014sta}. In terms of the composite effective vielbein
\begin{equation}
 e^{\cal A}_{\rm eff}{}_{\mu} =
  \alpha e^{\cal A}{}_{\mu} + \beta E^{\cal A}{}_{\mu}\,,
\label{eq:tetradeff-def}
\end{equation}
it is constructed as
\begin{equation}
 g^{\rm eff}_{\mu\nu} =
  \eta_{\cal AB} e^{\cal A}_{\rm eff}{}_{\mu} e^{\cal B}_{\rm eff}{}_{\nu}\,,
\label{eq:geff-def}
\end{equation}
with constants $\alpha$ and $\beta$. Hence, the action that we will be considering in this work is
\begin{equation}\label{action_MG_effcoupl_partial}
\mathcal{S} = \int \mathrm{d}^4x \big[ \frac{\mpl^2}{2} \det{e} \, R[e]+\mpl^2m^2\sum_n \beta_n{\cal L}_n  +\det{e_{\rm eff}}\,\mathcal{L}_\phi(e_{\rm eff},\phi)\big]\,,
\end{equation}
where $\phi$ represents the matter field. With the composite matter coupling, the partially constrained vielbein formulation results in physical consequences different from those in other formulations, as we shall see in the next section. For the specific homogeneous and isotropic background of interest we will be concentrating on solutions in which $Y^{\cal AB}$ is symmetric. Nonetheless, perturbations around it introduce the antisymmetric part of $Y^{\cal AB}$ in different ways in different formulations.


\section{Absence of the Boulware-Deser ghost}
\label{absence_ghost}

The Einstein-Hilbert kinetic term for the physical metric is invariant under two separate local Lorentz transformations,
\begin{equation}
 e^A_{\ \mu} \to \lambda^A_{\ B}e^B_{\ \mu}, \qquad
 E^A_{\ \mu} \to \Lambda^A_{\ B}E^B_{\ \mu}. 
\end{equation}
However, the inclusion of the graviton mass terms breaks the two independent copies of local Lorentz transformations to a diagonal combination. Hence the action depends on $\Lambda^{-1}\lambda$, while it is independent of the overall local Lorentz transformation. Without loss of generality it is thus possible to make use of the overall Lorentz transformation to set the fiducial vielbein $E^A_{\ \mu}$ to the ADM form (\ref{eqn:ADMvielbein}). Moreover, any general vielbein can be written in the boosted ADM form. We thus reparametrize the physical vielbein $e^A_{\ \mu}$ as (\ref{eqn:boosted-ADMvielbein}). Similarly, we reparametrize the effective vielbein $e^A_{\mathrm{eff}\mu}$ by the lapse $N_{\mathrm{eff}}$, the shift $N_{\mathrm{eff}}^i$, the spatial vielbein $e^I_{\mathrm{eff}j}$ and the boost parameter $b^{\mathrm{eff}}_I$ as
\begin{equation}
 e^A_{\mathrm{eff}\mu} = 
  \left(
   \begin{array}{cc}
    N_{\mathrm{eff}}\gamma_\mathrm{eff} + N_\mathrm{eff}^kb^{\mathrm{eff}}_Le^L_{\mathrm{eff}k} & b^{\mathrm{eff}}_Le^L_{\mathrm{eff}j}\\
    N_{\mathrm{eff}}b_{\mathrm{eff}}^I+N_{\mathrm{eff}}^ke^L_{\mathrm{eff}k}\left(\delta^I_L+\frac{b_{\mathrm{eff}}^Ib^{\mathrm{eff}}_L}{\gamma_{\mathrm{eff}}+1}\right) & 
     e^L_{\mathrm{eff}j}\left(\delta^I_L+\frac{b_{\mathrm{eff}}^Ib^{\mathrm{eff}}_L}{\gamma_{\mathrm{eff}}+1}\right)
    \end{array}
       \right),
\end{equation}
where $b_{\mathrm{eff}}^I=\delta^{IJ}b^{\mathrm{eff}}_J$ and $\gamma_{\mathrm{eff}}=\sqrt{1+b_{\mathrm{eff}}^Ib^{\mathrm{eff}}_I}$. In the following we will closely follow the argument presented in \cite{Hinterbichler:2015yaa} to show that $N_{\mathrm{eff}}$ and $N_{\mathrm{eff}}^i$ are linear in $N$, $N^i$, $M$ and $M^i$. Substitution of the above decompositions of the two vielbeins $e^A_{\ \mu}$, $E^I_{\ \mu}$ and the effective vielbein $e^A_{\mathrm{eff}\mu}$ into the relation $ e^{\cal A}_{\rm eff}{}_{\mu} = \alpha e^{\cal A}{}_{\mu} + \beta E^{\cal A}{}_{\mu} $ results in
\begin{eqnarray}
 b^{\mathrm{eff}}_Le^L_{\mathrm{eff}j} & = & \alpha b_Le^L_{\ j}, \nonumber\\
 e^L_{\mathrm{eff}j}\left(\delta^I_L+\frac{b_{\mathrm{eff}}^Ib^{\mathrm{eff}}_L}{\gamma_{\mathrm{eff}}+1}\right) & = & \alpha e^L_{\ j}\left(\delta^I_L+\frac{b^Ib_L}{\gamma+1}\right) + \beta E^I_{\ j}, \label{eqn:eeff1}
\end{eqnarray}
and
\begin{eqnarray}
    N_{\mathrm{eff}}\gamma_\mathrm{eff} + N_\mathrm{eff}^kb^{\mathrm{eff}}_Le^L_{\mathrm{eff}k} & = & \alpha (N\gamma + N^kb_Le^L_{\ k} )
     + \beta M, \nonumber\\
 N_{\mathrm{eff}}b_{\mathrm{eff}}^I+N_{\mathrm{eff}}^ke^L_{\mathrm{eff}k}\left(\delta^I_L+\frac{b_{\mathrm{eff}}^Ib^{\mathrm{eff}}_L}{\gamma_{\mathrm{eff}}+1}\right)
  & = & \alpha 
  \left[ Nb^I+N^ke^L_{\ k}\left(\delta^I_L+\frac{b^Ib_L}{\gamma+1}\right) \right]
  + \beta M^kE^I_{\ k}. \label{eqn:eeff2}
\end{eqnarray}
Since the right hand sides of (\ref{eqn:eeff1}) do not depend on $N$, $N^i$, $M$ and $M^i$, by solving these equations with respect to $e^I_{\mathrm{eff}\mu}$ and $b^{\mathrm{eff}}_I$, gives rise to expressions of $e^I_{\mathrm{eff}\mu}$ and $b^{\mathrm{eff}}_I$ in terms of $e^L_{\ \rho}$, $E^L_{\ \rho}$ and $b_L$. By substituting these solutions to (\ref{eqn:eeff2}), one obtains a set of linear equations for $N_{\mathrm{eff}}$ and $N_{\mathrm{eff}}^i$. The right hand sides of these linear equations are still linear in $N$, $N^i$, $M$ and $M^i$. Therefore, by solving them with respect to $N_{\mathrm{eff}}$ and $N_{\mathrm{eff}}^i$, one sees that $N_{\mathrm{eff}}$ and $N_{\mathrm{eff}}^i$ are linear in $N$, $N^i$, $M$ and $M^i$. Furthermore, the Hamiltonian density of the matter fields propagating on $g^{\mathrm{eff}}_{\mu\nu}$ is linear in $N_{\mathrm{eff}}$ and $N_{\mathrm{eff}}^i$. Thus, it is linear in $N$, $N^i$, $M$ and $M^i$. Needless to say, the Hamiltonian density of the matter fields propagating on $g_{\mu\nu}$ is linear in $N$ and $N^i$ and independent of $M$ and $M^i$. Also, each term in the graviton mass terms (\ref{eqn:def-gravitonmassterm}) includes only one factor of $e^A_{\ 0}$ or $E^A_{\ 0}$. Since $e^A_{\ 0}$ and $E^A_{\ 0}$ are linear in $N$, $N^i$, $M$ and $M^i$ and other components of vielbeins are independent of $N$, $N^i$, $M$ and $M^i$, the graviton mass terms are linear in $N$, $N^i$, $M$ and $M^i$. Therefore, the Hamiltonian of the system is linear in $N$, $N^i$, $M$ and $M^i$. Following \cite{Hinterbichler:2012cn}, one can thus show that there is a primary constraint that removes the Boulware-Deser ghost as follows: The equations of motion for $N^i$ ($i=1,2,3$) do not depend on $N$, $M$ and $M^j$ but depend on the boost parameters $b_L$ ($L=1,2,3$). Thus one can solve these $3$ equations with respect to the $3$ parameters $b_L$ and the solutions are independent of $N$, $N^i$, $M$ and $M^i$. Therefore, after substituting them, the Hamiltonian is independent of $N^i$ and linear in $N$, $M$ and $M^i$. The equation of motion for $N$ is thus a constraint. This is the primary constraint that removes the Boulware-Deser ghost, with the help of the associated secondary constraint.

In the case studied in \cite{Hinterbichler:2015yaa}, this argumentation unfortunately had a loophole. This loophole was due to the fact that the structure of the Hamiltonian drastically changes after solving the equations of motion for the spatial rotational part of the local Lorentz transformation. This fact was not appreciated in \cite{Hinterbichler:2015yaa} but was later noticed in \cite{deRham:2015cha}. In the metric formulation (or equivalently the constrained vielbein formulation) of the dRGT theory without the composite matter coupling, the (would-be) Boulware-Deser ghost is removed by a primary constraint and the associated secondary constraint in the phase space spanned by $g_{ij}$ and their canonical momenta. The presence of such a primary constraint is equivalent to the fact that the Hamiltonian density is linear in the lapse after integrating out the shift vector. The unconstrained vielbein formulation (after fixing the overall local Lorentz transformation), on the other hand, includes not only $g_{ij}$, the lapse and the shift but also the boost parameters and the spatial rotation parameters as basic variables. One thus needs to show that the Hamiltonian density is linear in the lapse after integrating out not only the shift but also the boost and spatial rotation parameters. This is indeed the case in dRGT theory without the composite matter coupling. In the presence of the composite matter coupling, the argument \cite{Hinterbichler:2015yaa} proves that the Hamiltonian density is linear in the lapse after integrating out the shift and the boost parameters. However, there still remain the spatial rotation parameters. However, as pointed out in \cite{deRham:2015cha}, the Hamiltonian density becomes highly non-linear in the lapse after integrating out the spatial rotation parameters as well. Thus the unconstrained vielbein formulation of the effective coupling reintroduces the Boulware-Deser ghost in analogy to the metric formulation. Conversely, in the case of partially constrained vielbein formulation that we study here, the argumentation of \cite{Hinterbichler:2015yaa} is solid. The spatial rotational part of the local Lorentz transformation is already fixed by the symmetric condition (\ref{eqn:YIJ=YJI}) at the level of the theory and thus there are no spatial rotation parameters in the set of basic variables. After using the equations of motion for the shift, we can integrate out the shift and the boost parameters all together, and the Hamiltonian density remains linear in the lapse as explained above. Hence the loophole of the argumentation in \cite{Hinterbichler:2015yaa} is removed by the symmetric condition (\ref{eqn:YIJ=YJI}), at the price of losing local Lorentz symmetry. Our setup illustrates a correct realization of the argument given in \cite{Hinterbichler:2015yaa}.


\section{Cosmological background evolution}\label{sec:background_evolution}

The no-go result for flat FLRW solutions in the standard formulation of massive gravity \cite{PhysRevD.84.124046} has motivated to consider non-minimal matter couplings. In the standard coupling the St\"uckelberg field equation of motion imposes a non-dynamical scale factor. Non-minimal matter coupling through a consistent composite effective metric softens the restrictions on the scale factor and one can easily construct 
exact FLRW solutions with flat reference metric \cite{deRham:2014naa}. In the vielbein formulation this effective composite metric corresponds to a linear effective vielbein built additively out of the two vielbeins. In the metric formulation this non-minimal coupling reintroduces the Boulware-Deser ghost, however, this does not prevent one to view this coupling as effective field theories below some cutoff scale. It was argued that the Boulware-Deser ghost might remain absent in the unconstrained vielbein formulation, but unfortunately the absence of the secondary constraints was proven very soon. The setup that we are using here corresponds to the correct realisation of the argument given in \cite{Hinterbichler:2015yaa}. In the previous section we have explicitly shown the ghost absence in the partially constrained vielbein formulation. This motivates us to investigate the cosmological implications of this model further. For this purpose, we shall assume a background with $Y_{\cal [AB]}=0$. Note that from the graviton mass terms one only obtains non-linear contributions in $Y_{\cal [AB]}$. Thus, in the absence of the matter coupling through the linear effective vielbein one consistent solution would be $Y_{\cal [AB]}=0$. Even in the presence of the matter coupling we shall concentrate on the background with symmetric $Y_{\cal AB}$ for the FLRW background. Let us consider the following homogeneous and isotropic co-diagonal backgrounds for the dynamical and fiducial metric. In other words both metrics have a FLRW form in the same coordinate system \footnote{
Although we use a generic FLRW form for the fiducial metric, the geometry of the fiducial space in massive gravity should be specified at a fundamental level. This connection with a fundamental geometry might then impose further restrictions on $N_f$ and $a_f$. For a list of FLRW charts of maximally symmetric space-times, see e.g. Ref.\cite{Gumrukcuoglu:2012wt}.
}

\begin{eqnarray}
ds_g^2&=&-N^2 dt^2 +a^2 \delta_{ij} dx^idx^j\,, \nonumber\\
ds_f^2&=& -N_f^2 dt^2 + a_f^2 \delta_{ij} dx^idx^j\,.
\end{eqnarray}
The corresponding tetrads read
\begin{equation}
e_\mu^0 = \delta_\mu^0 N\,,\qquad
e_\mu^I = \delta_\mu^I a\,,\qquad
E_\mu^0 = \delta_\mu^0 N_f\,,\qquad
E_\mu^I = \delta_\mu^I a_f\,,
\end{equation}
and similarly the diagonal symmetric matrix $Y_{\cal A B} $ takes the form
\begin{equation}
Y_{\cal A B} = {\rm diag}\left(-\frac{N}{N_f}\,,\;\frac{a}{a_f}\,,\;\frac{a}{a_f}\,,\;\frac{a}{a_f}\right)\,.
\label{eq:diagonalYbar}
\end{equation}
Accordingly, the composite effective metric on this background corresponds to the line element
\begin{equation}
ds^2_\eff = -N^2_\eff dt^2+a_\eff^2 \delta_{ij}dx^idx^j\,,
\end{equation}
where $N_{\eff}$ is the effective lapse function and $a_{\eff}$ stands for the effective scale factor
\begin{equation}
N_\eff \equiv \alpha\,N+\beta\,N_f \,,\qquad
a_\eff \equiv \alpha\,a+\beta\,a_f\,.
\end{equation}
As for the matter sector, we will only consider a generic scalar field $\phi$ minimally coupled to the effective metric
\begin{equation}
\mathcal{S} _\phi = \int \mathrm{d}^4x \det{e_{\rm eff}}\,\mathcal{L}_\phi(e_{\rm eff},\phi)=\int \mathrm{d}^4x \det{e_\eff}\,P\left( \chi, \phi  \right) \,,
\end{equation}
with $\chi=-\frac{1}{2}g^{\mu\nu}_\eff \partial_\mu\phi \partial_\nu\phi$. Since the scalar field has to be compatible with our above homogeneous and isotropic Ansaetze, we will impose that the scalar field depends only on time $\phi=\phi(t)$ at the background level. For later convenience let us also introduce the following quantities
\begin{eqnarray}\label{shortcuts}
&&H\equiv \frac{\dot{a}}{a\,N}\,, H_f\equiv \frac{\dot{a}_f}{a_f\,N_f} \,,\\
&&r\equiv \frac{N_f/a_f}{N/a}\,, \\
&&U(A)  \equiv  \beta_1 4A^3 + 6\beta_2 A^2 + 4\beta_3 A \,,
\end{eqnarray}
with $H$ denoting the expansion rate, $A\equiv a_0/a$ the ratio of the scale factors and $r$ the speed of light propagating in the fiducial metric with respect to the one propagating in the dynamical $g$ metric. The action on the homogeneous and isotropic background simplifies to
\begin{eqnarray}
\frac{S}{V} &=& \mpl^2\int dt \,a^3 N\,\Bigg\{-\Lambda-3H^2-m^2\beta_4-m^2\left[\rho_m+r A Q\right]\Bigg\} \nonumber\\
&&- m^2\mpl^2\int dt \,\beta_0 a_f^3 N_f+\int dt \,a_\eff^3 N_\eff P\left(  \chi, \phi\right)\,,
\label{eq:minisuperspace}
\end{eqnarray}
where we introduced the shortcut notations for manageability  
\begin{equation}
\rho_m (A)\equiv U(A)-\frac{A}{4}\, U_{,A} \,,\qquad
Q(A) \equiv \frac{1}{4}U_{,A}\,,
\end{equation}
with $U_{,A}=\partial_AU$.
After varying the mini-superspace action (\ref{eq:minisuperspace}) with respect to the lapse $N$ we obtain the Friedmann equation 
\begin{equation}
3\,H^2 = \Lambda + m^2 \beta_4 + m^2 \rho_m +\frac{\alpha\,a_\eff^3}{\mpl^2\,a^3}\rho\,,
\label{eq:eqN}
\end{equation}
where it becomes clear that $\rho_m$ denotes a dimensionless effective energy density from the mass term and $\rho \equiv 2\,\partial_{\chi}P(\chi,\phi)\,\chi - P(\chi,\phi)$ is the associated energy density of the matter field. The acceleration equation results from combining the equation of motion for the scale factor $a$ with the Friedmann equation
\begin{equation}
\frac{2\,\dot{H}}{N}= m^2\,J\,A\,(r-1) - \frac{\alpha\,a_\eff^3}{\mpl^2a^3}\left[\rho+\frac{a N_\eff}{a_\eff N}P
\right]\,,
\label{eq:eqa}
\end{equation}
with $J(A)\equiv \frac{1}{3}\,\partial_A \rho_m(A)$ and the pressure of the matter field $ P = P(\chi,\phi)$. Similarly, the variation of the mini-superspace action with respect to the scalar field yields 
\begin{equation}
\frac{1}{N_\eff}\,\dot{\rho}+3\,\frac{\dot{a}_\eff}{a_\eff N_\eff}\,(\rho+P)=0\,,
\label{eq:eqchi}
\end{equation}
which just correspond to the standard conservation equation for a field minimally coupled to the $g_\eff$ metric. Finally, the contracted Bianchi identity gives the constraint equation
\begin{equation}
\left( H-H_f A\right)\left[m^2\,\mpl^2J-\frac{\alpha\beta\,a_\eff^2}{a^2}P\right]=0  \,.
\label{eq:eqf}
\end{equation}
Generically, this equation defines two branches of solutions: Branch-I $H = H_f A$ and Branch-II $J \propto P$. When we study the perturbations, we will consider both branches of solutions separately.
At this stage, it is worth to mention that for Minkowski fiducial metric $a_f=a_0$ and $H_f=0$, only the second branch gives an expanding solution. In fact integrating the above equation for the case of Minkowski fiducial metric yields \cite{Gumrukcuoglu:2014xba}
\begin{equation}
m^2\mpl^2\,a^3\,Q+ \beta a_\eff^3\rho=a_0^3\,m^2 \mpl^2\kappa\,,
\label{eq:eqf-alt}
\end{equation}
where $\kappa$ is a dimensionless integration constant, independent of the normalization of the scale factor. The contracted Bianchi identity together with its integrated version in the case of Minkowski fiducial metric permit to relate the kinetic and potential energy density of the matter field to the graviton mass terms. The Friedmann equation can thus be brought into the form
\begin{equation}
3\,H^2=\Lambda + m^2 \beta_4 + m^2\left[\rho_m-\frac{\alpha}{\beta} \left(Q-\kappa\,A^3\right)\right]\,.
\label{eq:friedmann-final}
\end{equation}
Similarly, one can also remove the explicit $\chi$ dependence in the acceleration equation (\ref{eq:eqa}), yielding
\begin{equation}
\frac{2\,\dot{H}}{N} = -m^2\,\left[J\,A - \frac{\alpha}{\beta}\left(Q-\kappa\,A^3-J\right)
\right]\,.
\label{eq:acceleration-final}
\end{equation}
which is nothing else but the time derivative of the previous equation (\ref{eq:friedmann-final}). 
In the present study, we will not make use of this fact, and we will keep the fiducial metric arbitrary.

 As expected, the background evolution is exactly the same as in the metric formulation. Note that this is not surprising since we considered here background solutions with symmetric $Y_{\cal AB}$. There could be other interesting background evolution with a non-zero $Y_{\cal [AB]}$ background values. Even in our case with the vanishing $Y_{\cal [AB]}$ we expect non-trivial differences to the metric formulation at the level of the perturbations. 
 
 For later convenience in the study of perturbations we introduce the quantity
 \begin{eqnarray}
 \Gamma&\equiv& A\,J + \frac{(r-1)A^2}{2}\,\partial_A J\,,
\end{eqnarray}
 and the following relations that hold
\begin{eqnarray}
&&\frac{1}{N}\dot{\rho}_{m} = 3\,J\,A\,(H_f\,r\,A - H)\,,\nonumber\\
&&\frac{1}{N}\dot{J} = \frac{2\,(H_f\,r\,A - H)}{A\,(r-1)}\left(\Gamma-J\,A\right)\,,\nonumber\\
&&\frac{\dot{\Gamma}}{N} = 
(H_f\,r\,A - H)\left[(r-1)\rho_{m} + \frac{5\,r-3}{r-1}\,\Gamma -\frac{3\,r^2-2\,r+1}{r-1}\,J\,A\right]+\frac{\Gamma-J\,A}{(r-1)}\,\frac{\dot{r}}{N} \,.
\end{eqnarray}
%


\section{Stability of the perturbations}\label{sec:perturbations1}

As we have seen in the previous section, the coupling of the matter fields through a very specific admixture of the dynamical and fiducial vielbeins avoids the no-go result for flat FLRW solutions in massive gravity. In the metric formulation it was explicitly shown, that the Boulware-Deser ghost is absent around FLRW solutions \cite{deRham:2014naa}. In our partially constrained vielbein formulation this is true in general for any background. The ghost is maintained absent non-linearly by breaking explicitly the local Lorentz invariance. Even if the Boulware-Deser ghost is removed, it is important to investigate how the remaining physical degrees of freedom behave on top of the background that we are interested in. For this purpose we shall adopt 
the boosted ADM formulation where the physical tetrad can be written as
\begin{equation}
e^{\cal A}_{~~\mu} = \left(e^{-\omega}\right)^{\cal A}_{~~\cal B}\,\varepsilon^{\cal B}_{~~\mu}\,,
\end{equation}
where $\omega^{\cal A}_{~~\cal B}$ are the Lorentz transformations and $\varepsilon^{\cal B}_{~~\mu}$ corresponds to the ADM tetrad
\begin{equation}
\varepsilon^{\cal A}_{~~\mu} = \left(
\begin{array}{ll}
N(1+\Phi) & 0_j\\
\varepsilon^I_{~k}N^k & \varepsilon^I_{~j}
\end{array}
\right)\,,
\end{equation}
with the lapse perturbation $\Phi$. For the partially constrained tetrad formalism, only the boost part of the Lorentz transformations are kept, i.e. $\omega_{IJ}=0$. The perturbations of the boost parameters are decomposed into
\begin{equation}
\omega_{0I} = \partial_I v +v_I\,,
\label{eq:decompv}
\end{equation}
where $\delta^{IJ} \partial_J v_I=0$. The induced 3--metric can be built out of the triads $\varepsilon^I_{~j}$ as $h_{ij} = \varepsilon^I_{~i} \varepsilon^I_{~j}\delta_{IJ}$. The shift vector is decomposed as
\begin{equation}
N_i = h_{ij}N^j = a\,N\,(B_k+\partial_k B)\,,
\label{eq:decompN}
\end{equation}
while the triads are decomposed through
\begin{equation}
\varepsilon^I_{~i} = a\,(1+\,\psi)\delta^I_i+\frac{a\,\delta^{Ij}}{2}\left[\gamma_{ij}+\partial_{(i}E_{j)}+\left(\delta^k_i\delta^l_j-\frac{1}{3}\,\delta_{ij}\delta^{kl}\right)\partial_k\partial_lE\right]\,,
\label{eq:decompe}
\end{equation}
where $\delta^{ij}\partial_iB_j=\delta^{ij}\partial_iE_j=\delta^{ij}\partial_i\gamma_{jk} = \delta^{ij}\gamma_{ij}=0$.
Notice that the decomposition of the metric perturbations coincide with the ones in \cite{Gumrukcuoglu:2014xba} at linear order. 
Finally, the perturbation to the scalar field is introduced by
\begin{equation}
\phi = \phi_0+\delta\phi\,.
\label{eq:decompf}
\end{equation}
 In the following we study in detail the stability of tensor, vector and scalar perturbations in the two branches of solutions and exploit the
conditions to avoid ghost and Laplace instabilities. For this purpose we will further decompose the perturbations in Fourier modes with respect to the spatial coordinates
\begin{equation}
\mathcal{F}=\int \frac{\d^3k}{(2\pi)^{3/2}}\mathcal{F}_{\vec{k}}(t) \exp(i\vec{k}\cdot\vec{x}) +c.c. \, ,
\end{equation}
where $\mathcal{F}$ represents the perturbations ($\Phi$, $\psi$, $E$, $B$, $\delta\phi$, $v$) and ($B_I$, $E_I$, $v_I$, $h_{IJ}$) respectively. In the remainder of the paper, we will omit the subscript $\vec{k}$ associated with the mode functions for the sake of clarity of notation. We shall perform the stability analysis of the perturbations for each sector separately in what follows.

\subsection{Branch-I}
As we mentioned above from the constraint equation (\ref{eq:eqf}) one obtains two branches of solutions. Let us first investigate the perturbations on top of the Branch-I solutions with $H=AH_f$ and analyse the parameter space that is free from the ghost and Laplace instabilities.
\subsubsection{Tensor perturbations}
We start the stability analysis with the transverse-traceless fluctuations. Our action (\ref{action_MG_effcoupl_partial}) perturbed to second order in tensor perturbations has the form
\begin{equation}
 \mathcal{S}^{(2)}_{\rm tensor} = \frac{\mpl^2}{8}\int d^3k\,dt\,N\,a^3\,\left[\frac{1}{N^2}\dot{h}_{IJ}^\star \dot{h}^{IJ}-\left(\frac{k^2}{a^2}+m_{T}^2\right)h_{IJ}^\star h^{IJ} \right]\,,
\end{equation}
with the mass term defined as
\begin{equation}
m_{T}^2\equiv  m^2 \Gamma-\frac{\alpha\,\beta\,a_\eff N_\eff P\,A}{\mpl^2a\,N}\,.
\label{eq:tensormass-branch1}
\end{equation}
First of all the tensor perturbations have the right sign for the kinetic term and also the right condition for absence of gradient instabilities. The only condition comes from the requirement to avoid tachyonic instability $m_{T}>0$. 
\subsubsection{Vector perturbations}
As next we can compute our action perturbed to second order in vector perturbations:
\begin{eqnarray}
 \mathcal{S}^{(2)}_{\rm vector} &=& \frac{\mpl^2}{16}\int d^3k\,N\,dt\,k^2a^3 \left[
\frac{1}{N^2}\dot{E}_{I}^\star\dot{E}^I-\frac{2}{a\,N}\left(\dot{E}_{I}^\star B^I+B_{I}^\star \dot{E}^I\right)-m_T^2 E_{I}^\star E^I
\right.
\nonumber\\
&&\left.
\qquad\qquad\qquad
+\frac{4}{a^2}\,B_{I}^\star B^I-\frac{8}{k^2}\,m_{v}^2 \,v^\star_{I}v^I +\frac{8}{k^2}\,m_{vB}^2 \,\left(v_{I}^\star B^I+B_{I}^\star v^I\right)
\right]\,, \label{vect_bef1}
\end{eqnarray}
where we have introduced the following definitions for convenience 
\begin{eqnarray}
m_v^2 &&\equiv
m^2J\,A\,(r+1)+\frac{\alpha\,\beta\,A\,a_\eff^2}{\mpl^2a^2}\left[(r+1)\,\rho-\frac{a\,N_\eff}{a_\eff N}(\rho+P)\right]
\,, \nonumber \\
m_{vB}^2 &&\equiv m^2J\,A+\frac{\alpha\,\beta\,A\,a_\eff^2\rho}{a^2\mpl^2}\,.
\end{eqnarray}
Note that the vector fields $v_I$ and $B_I$ are non-dynamical degrees of freedom. We can therefore compute their equations of motion and integrate them out. Solving the equations of motion for $v_I$ and $B_I$, we obtain
\begin{equation}
v_{I} = \frac{m_{vB}^2}{m_v^2}\,B_{I} \,,\qquad
B_{I} = \frac{a}{2\,N}\left(1+\frac{2\,a^2}{k^2}\frac{m_{vB}^4}{m_v^2}\right)^{-1}\,\dot{E}_{I}\,.
\end{equation}
We can plug back these solutions in the quadratic action (\ref{vect_bef1}) and obtain
\begin{equation}
\mathcal{S}^{(2)}_{\rm vector} = \frac{\mpl^2}{16}\int d^3k\,N\,dt\,k^2a^3 \left[\mathcal{K}_V^2
\frac{1}{N^2}\dot{E}_{I}^\star\dot{E}^I-m_T^2 E_{I}^\star E^I 
\right]\,.
\end{equation}
with $\mathcal{K}_V^2= \left(1+\frac{k^2\,m_v^2}{2\,a^2\,m_{vB}^4}\right)^{-1}$.
We have to impose that the vector perturbations have the right sign for the kinetic term  $\mathcal{K}_V^2>0$ in order to avoid ghost instabilities. We can rewrite the kinetic term explicitly as
\begin{equation}
\mathcal{K}_V^2= \left[
1+\frac{k^2}{2\,A\,a^2}
\left(
\frac{r+1}{m^2J+\frac{\alpha\,\beta\,a_\eff^2\rho}{a^2\,\mpl^2}}-
\frac{a_\eff N_\eff}{a\,N\,\mpl^2}\,\frac{\alpha\,\beta\,(\rho+P)}{\left(m^2J+\frac{\alpha\,\beta\,a_\eff^2\rho}{a^2\mpl^2}\right)^2}
\right)
\right]^{-1}\,.
\end{equation}
At high momenta, the condition is rather cumbersome. At low momenta in the regime where $\mpl^2 m^2 J\gg \rho,P$, the kinetic term is positive if $m^2\,J>0$. In the other regime, where the matter is parametrically dominant over the mass terms, i.e. $\mpl^2 m^2 J\ll \rho,P$, 
we get the condition
\begin{equation}
\rho > \frac{P}{\frac{a_\eff N\,(r+1)}{a\,N_\eff}-1}\,.
\end{equation}
%
\subsubsection{Scalar perturbations}
As next we will study the stability of the scalar perturbations in our partially constrained vielbein formulation of the dRGT theory with the effective coupling. As we mentioned above, five dof appear in form of a scalar field $\Phi$, $B$, $v$, $\psi$, $E$, $\delta \phi$, from which three ($\Phi$, $B$ and $v$) are non-dynamical and will be integrated out. The equations of motion for $\Phi$, $B$ and $v$ are, respectively
\begin{eqnarray}
&&H\left(\frac{k^2B}{a}-3\,H\,\Phi+\frac{3\,\dot{\psi}}{N}\right)
+\frac{\alpha^2\,a_\eff^3N\,(\rho+P)}{2\,\mpl^2a^3N_\eff c_s^2}\,\Phi+\frac{3\,A}{2}\left(m^2J+\frac{\alpha\,\beta\,a_\eff^2\rho}{a^2\mpl^2}\right)\,\psi
\nonumber\\
&& \qquad\qquad
+\frac{k^2}{a^2}\left(\psi+\frac{k^2}{6}E\right)-\frac{\alpha\,a_\eff^3(\rho+P)}{2\,\mpl^2\,a^3 c_s^2\dot{\phi}_0}\,\delta\dot{\phi}
+\frac{\alpha\,a_\eff^3}{2\,\mpl^2a^3}\left(P_{,\phi}-P_{,\phi\chi}\,\frac{\dot{\phi}_0^2}{N_\eff^2}\right)\delta \phi
=0
\,,\nonumber\\
&& 
\frac{1}{N}\left(\dot{\psi}+\frac{k^2}{6}\,\dot{E}\right)-H\,\Phi
+\frac{\alpha\,a_\eff^2(\rho+P)}{2\,\mpl^2\,a^2}\,\frac{\delta\phi}{\dot{\phi}_0/N_\eff}
-\frac{a\,A}{2}\left(m^2J+\frac{\alpha\,\beta\,a_\eff^2\rho}{a^2\mpl^2}\right)v
=0\,,
\nonumber\\
&& \left(m^2J+\frac{\alpha\,\beta\,a_\eff^2\rho}{a^2\mpl^2}\right)B-
\left[(r+1)\left(m^2J+\frac{\alpha\,\beta\,a_\eff^2\rho}{a^2\mpl^2}\right)-\frac{\alpha\,\beta\,a_\eff N_\eff (\rho+P)}{\mpl^2a\,N}\right]\,v
\nonumber\\
&&\qquad\qquad\qquad\qquad\qquad\qquad\qquad\qquad\qquad\qquad
-\frac{\alpha\,\beta\,a_\eff (r-1)(\rho+P)}{\mpl^2a^2 \dot{\phi}_0/N_\eff}\,\delta\phi=0\,,
\label{eq:solPhiBv1}
\end{eqnarray}
where we made use of the sound speed of the matter field 
\begin{equation}
 c_s^2 \equiv \frac{P_{,\chi}}{\rho_{,\chi}}=\frac{\partial_{\chi}P(\chi,\phi)}{2\,\partial_{\chi}^2P(\chi,\phi)\,\chi+\partial_{\chi}P(\chi,\phi)}\,.
 \end{equation}
Using the solutions of these equations reduces the action to contain three degrees of freedom only. We find that a convenient basis to remove the would-be Boulware-Deser degree is
\begin{equation}
Y_{1} = \delta\phi - \frac{\alpha\,\dot{\phi}_0}{H\,N_\eff}\,\left(\psi+\frac{k^2}{6}\,E\right) \,,\qquad Y_{2} = \frac{1}{2}\,E\,.
\end{equation}
This choice of basis is just convenience since the subhorizon limit of the kinetic term turns out to be of order $k^0$. The basis used for Branch-II will be a different one. In this basis the mode $\psi$ becomes non-dynamical. Integrating it out, we end up with two dynamical modes, with action
\begin{equation}
 \mathcal{S}_{\rm scalar}^{(2)} = \frac{\mpl^2}{2}\int d^3k \,N\,dt\,a^3 \,\left[
\frac{\dot{Y}^\dagger}{N}\cdot K \cdot\frac{\dot{Y}}{N}
+\frac{\dot{Y}^\dagger}{N}\cdot M \cdot Y - Y^\dagger \cdot M \cdot \frac{\dot{Y}}{N} 
-Y^\dagger\cdot\Omega^2\cdot Y
\right]
\,.
\end{equation}
The exact form of the $2\times2$ matrices $K^T=K$, $(\Omega^2)^T=\Omega^2$ and $M^T=-M$ are very cumbersome, although the subhorizon expansion yields relatively manageable expressions. The components of the kinetic matrix at leading order in this expansion (${\cal O}(k^0)$) gives
\begin{eqnarray}
K_{11} &=& \frac{a_\eff^3N\,N_\eff(\rho+P)}{\mpl^2a^3c_s^2\,\dot{\phi}_0^2}+{\cal O}(k^{-1})
\,,\nonumber\\
K_{22} &=&\frac{3\,\left(m^2J+\frac{\alpha\,\beta\,a_\eff^2\rho}{a^2\mpl^2}\right)^2a^4A}{2}
\left[\frac{\left(m^2J+\frac{\alpha\,\beta\,a_\eff^2\rho}{a^2\mpl^2}\right)A
+\frac{\alpha^2a_\eff^2(\rho+P)}{a^2\mpl^2}-2\,H^2}{r\,\left(m^2J+\frac{\alpha\,\beta\,a_\eff^2\rho}{a^2\mpl^2} - \frac{\alpha\,\beta\,a_\eff^2(\rho+P)}{a^2\mpl^2}\right)} \right.\nonumber\\
&&\left.\qquad\qquad+\frac{\alpha^2a_\eff^3N\,A\,(\rho+P)}{6\,\mpl^2a^3c_s^2H^2N_\eff}\right]
+{\cal O}(k^{-1})
\,,\nonumber\\
K_{12}&=& \frac{\alpha\,a_\eff^3N\,A\,(\rho+P)}{2\,\mpl^2a\,c_s^2\,H\,\dot{\phi}_0}\,\left(m^2J+\frac{\alpha\,\beta\,a_\eff^2\rho}{a^2\mpl^2}\right)+{\cal O}(k^{-1})
\,.
\end{eqnarray}
One can rotate the basis such that the eigenvalues become
\begin{eqnarray}
\kappa_1 &=& K_{11} =
\frac{a_\eff^3N\,N_\eff(\rho+P)}{\mpl^2a^3c_s^2\,\dot{\phi}_0^2}+{\cal O}(k^{-1})
\,,\nonumber\\
\kappa_2 &=& \frac{\det K}{K_{11}} = 
\frac{3\,a^4\,A}{2\,r\,\beta}\left(m^2J+\frac{\alpha\,\beta\,a_\eff^2\rho}{a^2\mpl^2}\right)^2\,\left[
\frac{\tfrac{a_\eff}{a}\left(m^2J+\frac{\alpha\,\beta\,a_\eff^2\rho}{a^2\mpl^2}\right)-2\,\beta\,H^2}{\left(m^2J+\frac{\alpha\,\beta\,a_\eff^2\rho}{a^2\mpl^2} - \frac{\alpha\,\beta\,a_\eff^2(\rho+P)}{a^2\mpl^2}\right)}-\alpha
\right]
\,.
\end{eqnarray}
In the regime where the matter sector is parametrically subdominant under the mass term, the second eigenvalue gives a condition reminiscent of Higuchi bound, i.e. $m^2J(m^2J\,A-2\,H^2)/r>0$. In massive gravity, $H_f$ itself is determined by the geometry of the fiducial metric, which is a fundamental quantity. Therefore, keeping $H_f$ generic implicitly neglects some necessary information. For instance, if the fiducial metric is maximally symmetric, this gives a new relation between $J$ and the matter components. For this example, the second kinetic eigenvalue is simply
\begin{equation}
\kappa_2\Bigg\vert_{\dot{H}_f=0} = \frac{3\,\alpha\,a_\eff^3(\rho+P)\left(2\,\mpl^2H^2(r-1)+\frac{\alpha\,a_\eff^2N_\eff(\rho+P)}{a^2N}\right)^2}{2\,\mpl^4(r-1)^2\left(2\,\mpl^2H^2(r-1)+\frac{\alpha\,a_\eff^3(\rho+P)}{a^3}\right)}+{\cal O}(k^{-1})\,,
\end{equation}
which is positive if the sufficient conditions $\alpha>0$ and $r>1$ are satisfied. This example can also be applied to the vector mode and the vector kinetic term can be found to be positive if these conditions, along with $\beta>0$ are satisfied. The anti-symmetric mixing-matrix is of order $M_{12}={\cal O}(k^0)$, so it does not contribute to the speed of propagation. Finally, the components of the mass matrix are
\begin{eqnarray}
(\Omega^2)_{11} &=& \frac{a_\eff\,N_\eff^3(\rho+P)}{a^3\mpl^2N\,\dot{\phi}_0^2}\,k^2+{\cal O}(k)\,,\nonumber\\
(\Omega^2)_{22} &=& \left[-2\,H^2m_T^2
+\frac{A\left(m^2J+\frac{\alpha\,\beta\,a_\eff^2\rho}{a^2\mpl^2}\right)}{2}\left(4\,m_T^2+\frac{\alpha\,a_\eff^2\,N_\eff(\rho+P)}{\mpl^2a^2N}-2\,H^2\right)
\right.\nonumber\\
&& \left.+\frac{A^2\left(m^2J+\frac{\alpha\,\beta\,a_\eff^2\rho}{a^2\mpl^2}\right)^2}{4}\left(\frac{\alpha^2a_\eff N_\eff(\rho+P)}{\mpl^2a\,H^2N}-2\,r\right)\right]\,a^2k^2+{\cal O}(k)
\,,\nonumber\\
(\Omega^2)_{12} &=& \frac{\alpha\,a_\eff N_\eff^2 A\,(\rho+P)}{2\,\mpl^2a\,H\,N\,\dot{\phi}_0}\,\left(m^2J+\frac{\alpha\,\beta\,a_\eff^2\rho}{a^2\mpl^2}\right)k^2+{\cal O}(k)\,.
\end{eqnarray}
For a mode with subhorizon frequency $\omega = C_S\,\frac{k}{a}+{\cal O}(k^0)$, and taking into account the leading order expressions of the above matrices, we can obtain the propagation speeds of the modes $C_S^2$ by solving
\begin{equation}
\det\left[-\omega^2\,K+\Omega^2\right]=0\,,
\end{equation}
or
\begin{equation}
\frac{k^4}{a^4}\,\det[K]\,C_S^4 +\frac{k^2}{a^2}\,\left({\rm Tr}[\Omega^2\cdot K]-{\rm Tr}[\Omega^2]{\rm Tr}[K]\right)\,C_S^2+\det[\Omega^2]=0\,.
\end{equation}
%

%
\subsection{Branch-II}
We would like now study also the behaviour of perturbations on top of the Branch-II solutions with the constraint equation $J=\frac{\alpha\beta\,a_\eff^2}{m^2\,\mpl^2a^2}P$. Similarly as in the previous section we will perform the stability analysis of the perturbations for each sector separately. 
\subsubsection{Tensor perturbations}
We shall again start our stability analysis with the transverse-traceless part of fluctuations. 
The action (\ref{action_MG_effcoupl_partial}) perturbed to second order in tensor perturbations is this time
\begin{equation}
 \mathcal{S}^{(2)}_{\rm tensor} = \frac{\mpl^2}{8}\int d^3k\,dt\,N\,a^3\,\left[\frac{1}{N^2}\dot{h}_{IJ}^\star \dot{h}^{IJ}-\left(\frac{k^2}{a^2}+\bar{m}_{T}^2\right)h_{IJ}^\star h^{IJ} \right]\,,
\end{equation}
with the different mass term defined now as
\begin{equation}
\bar{m}_{T}^2\equiv  m^2\left( \Gamma-\frac{\alpha\,\beta\,a_\eff N_\eff P\,A}{\mpl^2a\,N} \right)\,.
\label{eq:tensormass-branch2}
\end{equation}
Again the tensor perturbations on top of the Branch-II solutions have already the right sign for the kinetic term and also for the absence of gradient instabilities. The condition for absence of tachyonic instability this time becomes $\bar{m}_{T}>0$. Actually, the tensor perturbations coincide exactly with the tensor perturbations obtained in \cite{Gumrukcuoglu:2014xba} for the metric formulation. This is no surprise, since the boost parameters $\omega_{\cal AB}$ contribute only to the vector and scalar perturbations (\ref{eq:decompv}) in our partially constrained formulation of the vielbein. Even if the mass term $\bar{m}_{T}^2$ has a very different form at first sight, it is actually the same as in Ref.\cite{Gumrukcuoglu:2014xba}. This becomes clear after using the following identification of the parameters between the two formulations:
\begin{equation}
\beta_1 \to -\frac{3}{2}(\alpha_3+4\,\alpha_4)\,,\quad
\beta_2 \to \alpha_2+3\,\alpha_3+6\,\alpha_4\,,\quad
\beta_3 \to -\frac{3}{2}(2\,\alpha_2+3\,\alpha_3+4\,\alpha_4)
\,.
\end{equation}
With this identification, we observe that $J$ does not get modified from one construction to the other, while 
\begin{equation}
\Gamma=\frac{A}{(A-1)^3}\left[(A-1)^2[(2\,r-1)A+r-2]J-(r-1)(Q+A^2\rho_{m})\right]\,.
\end{equation}
Using these relations, equation (\ref{eq:tensormass-branch2}) reduces in fact to the one in Ref.\cite{Gumrukcuoglu:2014xba}.

\subsubsection{Vector perturbations}
As in the previous section our vector modes consist of the non-dynamical degrees $B_I$, $v_I$ and the dynamical degrees $E_I$. The action quadratic in vector modes reads:
\begin{eqnarray}
 \mathcal{S}^{(2)}_{\rm vector} &=& \frac{\mpl^2}{16}\int d^3k\,N\,dt\,k^2a^3 \left[
\frac{1}{N^2}\dot{E}_{I}^\star\dot{E}^I-\frac{2}{a\,N}\left(\dot{E}_{I}^\star B^I+B_{I}^\star \dot{E}^I\right)-\bar{m}_T^2 E_{I}^\star E^I
\right.
\nonumber\\
&&\left.
\qquad\qquad\qquad
+\frac{4}{a^2}\,B_{I}^\star B^I-\frac{8}{k^2}\,\bar{m}_{v}^2 \,v^\star_{I}v^I +\frac{8}{k^2}\,\bar{m}_{vB}^2 \,\left(v_{I}^\star B^I+B_{I}^\star v^I\right)
\right]\,, \label{vect_bef}
\end{eqnarray}
where this time 
\begin{equation}
\bar{m}_v^2 \equiv \frac{\alpha\,\beta\,A\,a_\eff(\alpha\,r+\beta\,A)\,(\rho+P)}{a\,\mpl^2}\,,
\qquad
\bar{m}_{vB}^2 \equiv \frac{\alpha\,\beta\,A\,a_\eff^2\,(\rho+P)}{a^2\,\mpl^2}\,.
\end{equation}
We remark that the metric formulation corresponds to $v_I=B_I/(1+r)$, reducing to the action in \cite{Gumrukcuoglu:2014xba} at this stage. This can be verified by looking at the mass terms ${\cal L}_n$ quadratic in perturbations. In the boosted ADM formulation, one has
\begin{equation}
\left.\frac{\delta {\cal L}_n}{\delta v_I}\right\vert_{v_I= N_I/(1+r)} = 0\,.
\end{equation}
In other words, the fully constrained tetrad formalism, which is equivalent to the metric formulation, requires $\omega_{IJ}=0$ and $\omega_{0I} = N_I/(1+r)$ at linear order around FLRW. This is nothing but the symmetric vielbein condition, i.e. $Y_{\cal [AB]}=0$. We again depart from the metric formulation by solving the equations of motion for $v_I$, together with $B_I$, obtaining 
\begin{equation}
v_{I} = \frac{\bar{m}_{vB}^2}{\bar{m}_v^2}\,B_{I} \,,\qquad
B_{I} = \frac{a}{2\,N}\left(1+\frac{2\,a^2}{k^2}\frac{\bar{m}_{vB}^4}{\bar{m}_v^2}\right)^{-1}\,\dot{E}_{I}\,.
\end{equation}
Using these solutions back in the action, we obtain:
\begin{equation}
S^{(2)}_{\rm vector} = \frac{M_p^2}{16}\int d^3k\,N\,dt\,k^2a^3 \left[\bar{\mathcal{K}}_V^2
\frac{1}{N^2}\dot{E}_{I}^\star\dot{E}^I-\bar{m}_T^2 E_{I}^\star E^I 
\right]\,.
\end{equation}
The kinetic term for the vector mode can be written explicitly as
\begin{equation}
\bar{\mathcal{K}}_V^2\equiv \left(1+\frac{k^2\,\bar{m}_v^2}{2\,a^2\,\bar{m}_{vB}^4}\right)^{-1} = \left(
1+\frac{k^2\,\mpl^2 (\alpha\,r+\beta\,A)}{2\,\alpha\,\beta a^2\,A\,(\alpha+\beta\,A)^3\,(P+\rho)}
\right)^{-1}\,.
\end{equation}
At high momenta, assuming $\rho+P>0$, the no-ghost condition corresponds simply to $\alpha>0,\beta>0$.

\subsubsection{Scalar Perturbations}
As we introduced in the previous section the scalar sector contains six degrees of freedom in total: $\Phi$, $B$, $\psi$, $E$, $v$ and $\delta \phi$. Out of these $\Phi$, $B$, $v$ and a linear combination of the remaining ones are non-dynamical. Again, the metric formulation corresponds to fixing the boosts with $v=B/(1+r)$. The action, as usual, is not suitable for presentation. However, we quote here some of the intermediate steps. The equations of motion for $\Phi$, $B$ and $v$ are, respectively
\begin{eqnarray}
&&H\left(\frac{k^2B}{a}-3\,H\,\Phi+\frac{3\,\dot{\psi}}{N}\right)
+\frac{\alpha\,a_\eff^3(\rho+P)}{2\, \mpl^2a^3}\left(\frac{\alpha\,N\,\Phi}{c_s^2N_\eff}+\frac{3\,\beta\,a\,A\,\psi}{a_\eff}\right)
\nonumber\\
&& 
\qquad+\frac{\alpha\,a_\eff^3\,N_\eff}{2\, \mpl^2a^3\,\dot{\phi}_0}\,\left[\frac{2\,a\,(H-H_fr\,A)}{c_s^2\,a_\eff^2(r-1)\,A}\,\left(P\,A-\frac{m^2 \mpl^2 \,a\,N\,\Gamma}{\alpha\,\beta\,a_\eff N_\eff}\right)+3\,H_\eff(\rho+P)-\frac{P_{,\phi}\,\dot{\phi}_0}{c_s^2\,N_\eff}\right]\delta \phi
\nonumber\\
&&\qquad+\frac{k^2}{a^2}\left(\psi+\frac{k^2}{6}E\right)-\frac{\alpha\,a_\eff^3(\rho +P )}{2\, \mpl^2\,a^3 c_s^2\dot{\phi}_0}\,\delta\dot{\phi} =0
\,,\nonumber\\
&& 
\frac{1}{N}\left(\dot{\psi}+\frac{k^2}{6}\,\dot{E}\right)-H\,\Phi
+\frac{\alpha\,a_\eff^2(\rho +P )}{2\, \mpl^2\,a^2}\left(\frac{\delta\phi}{\dot{\phi}_0/N_\eff}-\beta\,a\,A\,v\right)=0\,,
\nonumber\\
&& a_\eff\,B-a\,(\alpha\,r+\beta\,A)\,v-\frac{(r-1)\delta\phi}{\dot{\phi}_0/N_\eff}=0\,.
\label{eq:solPhiBv2}
\end{eqnarray}
Using the solutions of these equations reduces the action to contain three degrees of freedom only. The convenient basis to remove the would-be Boulware-Deser ghost on top of Branch-II solutions is this time
\begin{equation}
Y_{1 }= \delta\phi - \frac{\alpha\,\dot{\phi}_0}{H\,N_\eff}\,\left(\psi+\frac{k^2}{6}\,E\right) \,,\qquad Y_{2} = \frac{k}{2}\,E\,.
\end{equation}
Once $\delta\phi$ and $E$ are expressed in terms of $Y_1$ and $Y_2$, the mode $\psi$ becomes non-dynamical. Integrating it out, we end up with two dynamical modes, with action
\begin{equation}
\mathcal{S}_{\rm scalar}^{(2)} = \frac{\mpl^2}{2}\int d^3k \,N\,dt\,a^3 \,\left[
\frac{\dot{Y}^\dagger}{N}\cdot \bar{K} \cdot\frac{\dot{Y}}{N}
+\frac{\dot{Y}^\dagger}{N}\cdot \bar{M} \cdot Y - Y^\dagger \cdot \bar{M} \cdot \frac{\dot{Y}}{N} 
-Y^\dagger\cdot\bar{\Omega}^2\cdot Y
\right]
\,,
\end{equation}
with $2\times2$ matrices $\bar{K}^T=\bar{K}$, $(\bar{\Omega}^2)^T=\bar{\Omega}^2$ and $\bar{M}^T=-\bar{M}$.
\begin{eqnarray}
\bar{K}_{12}&=& 
-\frac{2\,k\,N_\eff\,H}{\alpha\,\dot{\phi}_0}\left[
1-\frac{c_s^2\left[k^2(\alpha\,r+\beta\,A)-3\,\beta\,a^2\,A\,\tfrac{\dot{H}}{N}\right]}{\alpha\,\beta\,a^2A\tfrac{\dot{H}^2}{H^2\,N\,N_\eff}}\,
\mathcal{G}
\right]^{-1}
\,,\nonumber\\
\bar{K}_{11} &=& -\frac{\bar{K}_{12}\,H}{k\,\alpha\,\beta\,a^2A\,\tfrac{\dot{\phi}_0}{N_\eff}\,\tfrac{\dot{H}}{N}}\,\left[k^2(\alpha\,r+\beta\,A)-3\,\beta\,a^2\,A\,\tfrac{\dot{H}}{N}\right]\,,\nonumber\\
\bar{K}_{22} &=&\frac{\bar{K}_{12}}{\bar{K}_{11}}\left[\bar{K}_{12}+\frac{2\,k\,H}{\alpha\,\tfrac{\dot{\phi}_0}{N_\eff}}\right]
=-\frac{k\,\beta\,a^2\,A\,\tfrac{\dot{H}}{N} \left(2\,k\,H+\bar{K}_{12}\,\alpha\,\frac{\dot{\phi}_0}{N_\eff}\right)}{H\left[k^2(\alpha\,r+\beta\,A)-3\,\beta\,a^2\,A\,\tfrac{\dot{H}}{N}\right]}\,
\,.
\end{eqnarray}
where we have defined
\begin{equation}
\mathcal{G}=\left(1+\frac{3\,\alpha\,\beta\,a^2\,c_s^2(H-H_fA)(r-1)\,A\,\tfrac{\dot{H}}{N}}{a_\eff^2(\beta\,H+\alpha\,H_f)\,\bar{m}_T^2}\right)^{-1}.
\end{equation}
One can rotate the basis such that the eigenvalues are:
\begin{eqnarray}
\bar{\kappa}_1 &=& \bar{K}_{11} =
\frac{2\,\left(\frac{k^2}{a^2}(\alpha\,r+\beta\,A)-3\,\beta\,A\,\tfrac{\dot{H}}{N}\right)}{\alpha^2\beta\,A\,\left(\tfrac{\dot{H}}{N\,H^2}\right)\left(\tfrac{\dot{\phi}_0}{N_\eff}\right)^2}
\left[
1-\frac{c_s^2\left[k^2(\alpha\,r+\beta\,A)-3\,\beta\,a^2\,A\,\tfrac{\dot{H}}{N}\right]}{\alpha\,\beta\,a^2A\tfrac{\dot{H}^2}{H^2\,N\,N_\eff}}
\mathcal{G}
\right]^{-1}
\,,\nonumber\\
\bar{\kappa}_2 &=& \frac{\det \bar{K}}{\bar{K}_{11}} = 2\,k^2\,\left[3-\frac{k^2}{a^2}\,\frac{N}{\dot{H}\,\beta\,A}\,(\alpha\,r+\beta\,A)\right]^{-1}\,.
\end{eqnarray}
The full no-ghost conditions are quite opaque, so we consider the subhorizon limit of the action. In this limit the kinetic matrix is diagonal, with $\bar{K}_{12} = {\cal O}(k^{-1})$ and
\begin{eqnarray}
\bar{K}_{11} &=& \frac{a_\eff^3(\rho+P)}{M_p^2a^3c_s^2\,\tfrac{\dot{\phi}_0^2}{N\,N_\eff}}\,
\left[
1+\frac{3\,a_\eff\,\alpha^2\beta^2c_s^2(1-r)\,A}{2\,a}\,\frac{H-H_fA}{\beta\,H+\alpha\,H_f}\,\frac{\rho+P}{\mpl^2\bar{m}_T^2}
\right] + {\cal O}(k^{-1})\,,\nonumber\\
\bar{K}_{22} &=& \frac{\alpha\,\beta\,a_\eff^3A\,(\rho+P)}{\mpl^2a(\alpha\,r+\beta\,A)}+ {\cal O}(k^{-1})\,.
\end{eqnarray}
The mixing matrix already has one independent component, which is, at leading order:
\begin{equation}
\bar{M}_{12} = \frac{\alpha\,\beta\,a_\eff^3(H\,r-H_fA)(\rho+P)}{2\,\mpl^2\,a^3(\beta\,H+\alpha\,H_f)(\alpha\,r+\beta\,A)\tfrac{\dot{\phi}_0}{N_\eff}}\,k+{\cal O}(k^0)\,.
\end{equation}
Finally, the mass matrix is also diagonal at leading order, with
\begin{eqnarray}
(\bar{\Omega}^2)_{11} &=& \frac{a_\eff^3r\,(\rho+P)}{a^5M_p^2(\alpha\,r+\beta\,A)\tfrac{\dot{\phi}_0^2}{N_\eff^2}}\,k^2+{\cal O}(k)\,,\nonumber\\
(\bar{\Omega}^2)_{22} &=& \frac{2\,\bar{m}_T^2(H-H_f\,A)\,N_\eff}{3\,N\,A(r-1)(\beta\,H+\alpha\,H_f)}
\,k^2+{\cal O}(k)\,.
\end{eqnarray}

Then, for a mode with subhorizon frequency $\omega^2 = C_S^2\,\frac{k^2}{a^2}+{\cal O}(k)$, we can obtain the propagation speeds of the modes $C_S^2$ by solving
\begin{equation}
\det\left[-\omega^2\,\bar{K}+\left(\frac{1}{N}\dot{\bar{K}}+3\,H\,\bar{K}+2\,\bar{M}\right)(-i\,\omega)+\left(3\,H\,\bar{M}+\frac{1}{N}\dot{\bar{M}}+\bar{\Omega}^2\right)\right]=0\,.
\end{equation}
Considering the leading order terms of the matrices, the equation can be simplified as
\begin{equation}
\det\left[-\omega^2\,\bar{K}+2\,\bar{M}(-i\,\omega)+\bar{\Omega}^2\right]=0\,,
\end{equation}
or
\begin{equation}
\frac{k^4}{a^4}\,\bar{K}_{11}\bar{K}_{22}\,C_S^4 -\frac{k^2}{a^2}\,\left(4\,\bar{M}_{12}^2+\bar{K}_{11}\bar{\Omega}^2_{22}+\bar{K}_{22}\bar{\Omega}^2_{11}\right)C_S^2+\bar{\Omega}^2_{11}\bar{\Omega}^2_{22}=0
\end{equation}
This is a quadratic equation for $C_S^2$ and the solutions can be easily found, although they are not very instructive. An interesting limit is $\beta\to0$, which corresponds to the minimally-coupled matter limit:
\begin{equation}
 C_{S,1}^2= c_s^2+{\cal O}(\beta)\,,\qquad
C_{S,2}^2 = \frac{2\,\mpl^2\,r\,(H-H_fA)\bar{m}_T^2}{3\,\alpha^3\,H_f (r-1)\,A^2(\rho+P)\,\beta}+{\cal O}(\beta^0)\,.
\end{equation}
In other words, for vanishing $\beta$ one of the modes becomes instantaneous, while the other one is the matter degree as it propagates with the sound speed of the fluid. This is the standard ``vanishing kinetic terms'' issue of dRGT in self-accelerating branch with minimally coupled matter \cite{Gumrukcuoglu:2011zh}.

\section{Conclusions}
\label{sec:conclusion}

The question of consistent matter couplings within the framework of massive gravity is crucial for  maintaining the ghost freedom. The specific quantum properties of the theory have motivated the consideration of an effective matter coupling \cite{deRham:2013qqa,deRham:2014naa}. 
This matter coupling through a very specific admixture of the dynamical and fiducial metric avoids further the No-go result for flat FLRW solutions. However, this coupling reintroduces the Boulware-Deser ghost. It is an indispensable question to address at what scale this ghostly degree of freedom enters. Of course these interactions can be still considered as a consistent effective field theory with the cut-off scale dictated by the mass of the ghost. On top of a FLRW background the corresponding Boulware-Deser ghost remains absent at second order of perturbations at the level of the action and its mass on this background is infinite. A crucial step would be to study the perturbations on top of an anisotropic background and investigate in detail the mass of the ghost.\\

Even if one can use the matter coupling through the composite effective metric in the sense of effective field theory, it would be more appealing to find consistent matter coupling in which the absence of the Boulware-Deser ghost is realised fully non-linearly. It was argued that this might indeed be possible in the unconstrained vielbein formulation of the theory \cite{Hinterbichler:2015yaa}. Using the boosted ADM decomposition of the vielbein one can easily show that the effective lapse and shift functions remain linear in the lapses and shifts of the two vielbeins after integrating out the boost parameters resulting in first class primary constraints. With the help of the associated secondary constraints the Boulware-Deser ghost can be eliminated. Unfortunately, the absence of the secondary constraints was shown very soon \cite{deRham:2015cha}. In other words the integration of the rotations gives rise to a highly non-linear dependence on the lapses. This means that also in the unconstrained vielbein formulation it is not possible to avoid consistently the Boulware-Deser ghost.\\

In this present paper we exhausted yet another formulation of the theory, the so called partially constrained vielbein formulation (along the lines of \cite{DeFelice:2015hla}), with the attempt to remove the Boulware-Deser ghost fully non-linearly. This formulation guarantees the absence of the ghost to all orders beyond the decoupling limit. For this purpose we also adapted to the boosted ADM decomposition.
After proving the ghost absence, we studied the cosmological application of this formulation within the context of the non-minimal matter coupling in form of a k-essence model. Even if the background evolution is the same as in the metric formulation for vanishing background boost parameters, the dynamics of the perturbations is crucially different. This allowed us to put different constraints on the parameters of the theory coming from the requirement of ghost and Laplacian instabilities absence.
An important question to further investigate would be whether or not the ghost absence can be maintained in a different decomposition of the vielbein rather than the boosted ADM decomposition.
It would be also very interesting to apply the partially constrained vielbein formulation to the new quasi dilaton extension of massive gravity or similarly to bigravity and investigate their cosmological implications.

\acknowledgments

AEG acknowledges financial support from the European Research
Council under the European Union's Seventh Framework Programme
(FP7/2007-2013) / ERC Grant Agreement n. 306425 ``Challenging General
Relativity''.  
This work was supported in part by WPI Initiative, MEXT, Japan and
Grant-in-Aid for Scientific Research 24540256. 
Part of the work has been done within the Labex ILP (reference
ANR-10-LABX-63) part of the Idex SUPER, and received financial state
aid managed by the Agence Nationale de la Recherche, as part of the
programme Investissements d'avenir under the reference
ANR-11-IDEX-0004-02. He is thankful to Institut Astrophysique de Paris
for warm hospitality.

\appendix

	\bibliographystyle{JHEPmodplain}
	\bibliography{partially_constrained}

\providecommand{\href}[2]{#2}\begingroup\raggedright\begin{thebibliography}{10}

\bibitem{Fierz:1939ix}
M.~Fierz and W.~Pauli, {\it {On relativistic wave equations for particles of
  arbitrary spin in an electromagnetic field}},  {\sl Proc.Roy.Soc.Lond.} {\bf
  A173} (1939) 211--232.

\bibitem{vanDam:1970vg}
H.~van Dam and M.~Veltman, {\it {Massive and massless Yang-Mills and
  gravitational fields}},  {\sl Nucl.Phys.} {\bf B22} (1970) 397--411,
  [\href{http://dx.doi.org/10.1016/0550-3213(70)90416-5}{{\sf
  doi:10.1016/0550-3213(70)90416-5}}].

\bibitem{Zakharov:1970cc}
V.~Zakharov, {\it {Linearized gravitation theory and the graviton mass}},  {\sl
  JETP Lett.} {\bf 12} (1970) 312.

\bibitem{deRham:2010ik}
C.~de~Rham and G.~Gabadadze, {\it {Generalization of the Fierz-Pauli action}},
  {\sl Phys.Rev.} {\bf D82} (2010) 044020,
  [\href{http://arxiv.org/abs/1007.0443}{{\sf arXiv:1007.0443}}].

\bibitem{deRham:2010kj}
C.~de~Rham, G.~Gabadadze, and A.~J. Tolley, {\it {Resummation of massive
  gravity}},  {\sl Phys.Rev.Lett.} {\bf 106} (2011) 231101,
  [\href{http://arxiv.org/abs/1011.1232}{{\sf arXiv:1011.1232}}].

\bibitem{Hassan:2011vm}
S.~Hassan and R.~A. Rosen, {\it {On Non-Linear Actions for Massive Gravity}},
  {\sl JHEP} {\bf 1107} (2011) 009, [\href{http://arxiv.org/abs/1103.6055}{{\sf
  arXiv:1103.6055}}], [\href{http://dx.doi.org/10.1007/JHEP07(2011)009}{{\sf
  doi:10.1007/JHEP07(2011)009}}].

\bibitem{Hassan:2011hr}
S.~Hassan and R.~A. Rosen, {\it {Resolving the ghost problem in non-linear
  massive gravity}},  {\sl Phys.Rev.Lett.} {\bf 108} (2012) 041101,
  [\href{http://arxiv.org/abs/1106.3344}{{\sf arXiv:1106.3344}}].

\bibitem{PhysRevD.84.124046}
G.~D'Amico, C.~de~Rham, S.~Dubovsky, G.~Gabadadze, D.~Pirtskhalava, and A.~J.
  Tolley, {\it Massive cosmologies},  {\sl Phys. Rev. D} {\bf 84} (Dec, 2011)
  124046, [\href{http://arxiv.org/abs/1108.5231}{{\sf 1108.5231}}].

\bibitem{Gumrukcuoglu:2011ew}
A.~E. Gumrukcuoglu, C.~Lin, and S.~Mukohyama, {\it {Open FRW universes and
  self-acceleration from nonlinear massive gravity}},  {\sl JCAP} {\bf 1111}
  (2011) 030, [\href{http://arxiv.org/abs/1109.3845}{{\sf arXiv:1109.3845}}].

\bibitem{PhysRevLett.109.171101}
A.~De~Felice, A.~E. G\"umr\"uk\ifmmode \mbox{\c{c}}\else
  \c{c}\fi{}\"uo\ifmmode~\breve{g}\else \u{g}\fi{}lu, and S.~Mukohyama, {\it
  Massive gravity: Nonlinear instability of a homogeneous and isotropic
  universe},  {\sl Phys. Rev. Lett.} {\bf 109} (Oct, 2012) 171101,
  [\href{http://arxiv.org/abs/1206.2080}{{\sf arXiv:1206.2080}}],
  [\href{http://dx.doi.org/10.1103/PhysRevLett.109.171101}{{\sf
  doi:10.1103/PhysRevLett.109.171101}}].

\bibitem{DeFelice:2013awa}
A.~De~Felice, A.~E. G{\"u}mr{\"u}k{\c c}{\"u}o{\u g}lu, C.~Lin, and
  S.~Mukohyama, {\it {Nonlinear stability of cosmological solutions in massive
  gravity}},  {\sl JCAP} {\bf 1305} (2013) 035,
  [\href{http://arxiv.org/abs/1303.4154}{{\sf arXiv:1303.4154}}],
  [\href{http://dx.doi.org/10.1088/1475-7516/2013/05/035}{{\sf
  doi:10.1088/1475-7516/2013/05/035}}].

\bibitem{Fasiello:2012rw}
M.~Fasiello and A.~J. Tolley, {\it {Cosmological perturbations in Massive
  Gravity and the Higuchi bound}},  {\sl J. Cosm. Astropart.} {\bf 1211} (2012)
  035, [\href{http://arxiv.org/abs/1206.3852}{{\sf arXiv:1206.3852}}],
  [\href{http://dx.doi.org/10.1088/1475-7516/2012/11/035}{{\sf
  doi:10.1088/1475-7516/2012/11/035}}].

\bibitem{Langlois:2012hk}
D.~Langlois and A.~Naruko, {\it {Cosmological solutions of massive gravity on
  de Sitter}},  {\sl Class.Quant.Grav.} {\bf 29} (2012) 202001,
  [\href{http://arxiv.org/abs/1206.6810}{{\sf arXiv:1206.6810}}],
  [\href{http://dx.doi.org/10.1088/0264-9381/29/20/202001}{{\sf
  doi:10.1088/0264-9381/29/20/202001}}].

\bibitem{MartinMoruno:2013gq}
P.~Martin-Moruno and M.~Visser, {\it {Is there vacuum when there is mass?
  Vacuum and non-vacuum solutions for massive gravity}},  {\sl Class. Quant.
  Grav.} {\bf 30} (2013) 155021, [\href{http://arxiv.org/abs/1301.2334}{{\sf
  arXiv:1301.2334}}],
  [\href{http://dx.doi.org/10.1088/0264-9381/30/15/155021}{{\sf
  doi:10.1088/0264-9381/30/15/155021}}].

\bibitem{Gumrukcuoglu:2011zh}
A.~E. Gumrukcuoglu, C.~Lin, and S.~Mukohyama, {\it {Cosmological perturbations
  of self-accelerating universe in nonlinear massive gravity}},  {\sl J. Cosm.
  Astropart.} {\bf 1203} (2012) 006,
  [\href{http://arxiv.org/abs/1111.4107}{{\sf arXiv:1111.4107}}],
  [\href{http://dx.doi.org/10.1088/1475-7516/2012/03/006}{{\sf
  doi:10.1088/1475-7516/2012/03/006}}].

\bibitem{Gumrukcuoglu:2012wt}
A.~E. Gumrukcuoglu, S.~Kuroyanagi, C.~Lin, S.~Mukohyama, and N.~Tanahashi, {\it
  {Gravitational wave signal from massive gravity}},  {\sl Class. Quant. Grav.}
  {\bf 29} (2012) 235026, [\href{http://arxiv.org/abs/1208.5975}{{\sf
  arXiv:1208.5975}}],
  [\href{http://dx.doi.org/10.1088/0264-9381/29/23/235026}{{\sf
  doi:10.1088/0264-9381/29/23/235026}}].

\bibitem{Huang:2012pe}
Q.-G. Huang, Y.-S. Piao, and S.-Y. Zhou, {\it {Mass-Varying Massive Gravity}},
  {\sl Phys.Rev.} {\bf D86} (2012) 124014,
  [\href{http://arxiv.org/abs/1206.5678}{{\sf arXiv:1206.5678}}],
  [\href{http://dx.doi.org/10.1103/PhysRevD.86.124014}{{\sf
  doi:10.1103/PhysRevD.86.124014}}].

\bibitem{PhysRevD.87.064037}
G.~D'Amico, G.~Gabadadze, L.~Hui, and D.~Pirtskhalava, {\it Quasidilaton:
  Theory and cosmology},  {\sl Phys. Rev. D} {\bf 87} (Mar, 2013) 064037,
  [\href{http://dx.doi.org/10.1103/PhysRevD.87.064037}{{\sf
  doi:10.1103/PhysRevD.87.064037}}].

\bibitem{DeFelice:2013tsa}
A.~De~Felice and S.~Mukohyama, {\it {Towards consistent extension of
  quasidilaton massive gravity}},  {\sl Phys.Lett.} {\bf B728} (2014) 622--625,
  [\href{http://arxiv.org/abs/1306.5502}{{\sf arXiv:1306.5502}}],
  [\href{http://dx.doi.org/10.1016/j.physletb.2013.12.041}{{\sf
  doi:10.1016/j.physletb.2013.12.041}}].

\bibitem{Khosravi:2011zi}
N.~Khosravi, N.~Rahmanpour, H.~R. Sepangi, and S.~Shahidi, {\it {Multi-Metric
  Gravity via Massive Gravity}},  {\sl Phys.Rev.} {\bf D85} (2012) 024049,
  [\href{http://arxiv.org/abs/1111.5346}{{\sf arXiv:1111.5346}}],
  [\href{http://dx.doi.org/10.1103/PhysRevD.85.024049}{{\sf
  doi:10.1103/PhysRevD.85.024049}}].

\bibitem{Akrami:2012vf}
Y.~Akrami, T.~S. Koivisto, and M.~Sandstad, {\it {Accelerated expansion from
  ghost-free bigravity: a statistical analysis with improved generality}},
  {\sl JHEP} {\bf 1303} (2013) 099, [\href{http://arxiv.org/abs/1209.0457}{{\sf
  arXiv:1209.0457}}], [\href{http://dx.doi.org/10.1007/JHEP03(2013)099}{{\sf
  doi:10.1007/JHEP03(2013)099}}].

\bibitem{Akrami:2013ffa}
Y.~Akrami, T.~S. Koivisto, D.~F. Mota, and M.~Sandstad, {\it {Bimetric gravity
  doubly coupled to matter: theory and cosmological implications}},  {\sl JCAP}
  {\bf 1310} (2013) 046, [\href{http://arxiv.org/abs/1306.0004}{{\sf
  arXiv:1306.0004}}],
  [\href{http://dx.doi.org/10.1088/1475-7516/2013/10/046}{{\sf
  doi:10.1088/1475-7516/2013/10/046}}].

\bibitem{Tamanini:2013xia}
N.~Tamanini, E.~N. Saridakis, and T.~S. Koivisto, {\it {The Cosmology of
  Interacting Spin-2 Fields}},  {\sl JCAP} {\bf 1402} (2014) 015,
  [\href{http://arxiv.org/abs/1307.5984}{{\sf arXiv:1307.5984}}],
  [\href{http://dx.doi.org/10.1088/1475-7516/2014/02/015}{{\sf
  doi:10.1088/1475-7516/2014/02/015}}].

\bibitem{Akrami:2014lja}
Y.~Akrami, T.~S. Koivisto, and A.~R. Solomon, {\it {The nature of spacetime in
  bigravity: two metrics or none?}},
  \href{http://arxiv.org/abs/1404.0006}{{\sf arXiv:1404.0006}}.

\bibitem{deRham:2014naa}
C.~de~Rham, L.~Heisenberg, and R.~H. Ribeiro, {\it {On couplings to matter in
  massive (bi-)gravity}},  {\sl Class.Quant.Grav.} {\bf 32} (2015) 035022,
  [\href{http://arxiv.org/abs/1408.1678}{{\sf arXiv:1408.1678}}],
  [\href{http://dx.doi.org/10.1088/0264-9381/32/3/035022}{{\sf
  doi:10.1088/0264-9381/32/3/035022}}].

\bibitem{deRham:2014fha}
C.~de~Rham, L.~Heisenberg, and R.~H. Ribeiro, {\it {Ghosts and Matter Couplings
  in Massive (bi-and multi-)Gravity}},  {\sl Phys.Rev.} {\bf D90} (2014)
  124042, [\href{http://arxiv.org/abs/1409.3834}{{\sf arXiv:1409.3834}}],
  [\href{http://dx.doi.org/10.1103/PhysRevD.90.124042}{{\sf
  doi:10.1103/PhysRevD.90.124042}}].

\bibitem{Yamashita:2014fga}
Y.~Yamashita, A.~De~Felice, and T.~Tanaka, {\it {Appearance of Boulware--Deser
  ghost in bigravity with doubly coupled matter}},  {\sl Int.J.Mod.Phys.} {\bf
  D23} (2014) 1443003, [\href{http://arxiv.org/abs/1408.0487}{{\sf
  arXiv:1408.0487}}], [\href{http://dx.doi.org/10.1142/S0218271814430032}{{\sf
  doi:10.1142/S0218271814430032}}].

\bibitem{Noller:2014sta}
J.~Noller and S.~Melville, {\it {The coupling to matter in Massive, Bi- and
  Multi-Gravity}},  \href{http://arxiv.org/abs/1408.5131}{{\sf
  arXiv:1408.5131}}.

\bibitem{Schmidt-May:2014xla}
A.~Schmidt-May, {\it {Mass eigenstates in bimetric theory with matter
  coupling}},  {\sl JCAP} {\bf 1501} (2015) 039,
  [\href{http://arxiv.org/abs/1409.3146}{{\sf arXiv:1409.3146}}],
  [\href{http://dx.doi.org/10.1088/1475-7516/2015/01/039}{{\sf
  doi:10.1088/1475-7516/2015/01/039}}].

\bibitem{Enander:2014xga}
J.~Enander, A.~R. Solomon, Y.~Akrami, and E.~Mortsell, {\it {Cosmic expansion
  histories in massive bigravity with symmetric matter coupling}},
  \href{http://arxiv.org/abs/1409.2860}{{\sf arXiv:1409.2860}}.

\bibitem{Solomon:2014iwa}
A.~R. Solomon, J.~Enander, Y.~Akrami, T.~S. Koivisto, F.~K{\"o}nnig, {\em
  et~al.}, {\it {Does massive gravity have viable cosmologies?}},
  \href{http://arxiv.org/abs/1409.8300}{{\sf arXiv:1409.8300}}.

\bibitem{Soloviev:2014eea}
V.~O. Soloviev, {\it {Bigravity in tetrad Hamiltonian formalism and matter
  couplings}},  \href{http://arxiv.org/abs/1410.0048}{{\sf arXiv:1410.0048}}.

\bibitem{Heisenberg:2014rka}
L.~Heisenberg, {\it {Quantum corrections in massive bigravity and new effective
  composite metrics}},  {\sl Class.Quant.Grav.} {\bf 32} (2015), no.~10 105011,
  [\href{http://arxiv.org/abs/1410.4239}{{\sf arXiv:1410.4239}}],
  [\href{http://dx.doi.org/10.1088/0264-9381/32/10/105011}{{\sf
  doi:10.1088/0264-9381/32/10/105011}}].

\bibitem{Huang:2015yga}
Q.-G. Huang, R.~H. Ribeiro, Y.-H. Xing, K.-C. Zhang, and S.-Y. Zhou, {\it {On
  the uniqueness of the non-minimal matter coupling in massive gravity and
  bigravity}},  {\sl Phys. Lett.} {\bf B748} (2015) 356--360,
  [\href{http://arxiv.org/abs/1505.02616}{{\sf arXiv:1505.02616}}],
  [\href{http://dx.doi.org/10.1016/j.physletb.2015.07.003}{{\sf
  doi:10.1016/j.physletb.2015.07.003}}].

\bibitem{Heisenberg:2015wja}
L.~Heisenberg, {\it {Non-minimal derivative couplings of the composite
  metric}},  \href{http://arxiv.org/abs/1506.00580}{{\sf arXiv:1506.00580}}.

\bibitem{Blanchet:2015sra}
L.~Blanchet and L.~Heisenberg, {\it {Dark Matter via Massive (bi-)Gravity}},
  {\sl Phys. Rev.} {\bf D91} (2015) 103518,
  [\href{http://arxiv.org/abs/1504.00870}{{\sf arXiv:1504.00870}}],
  [\href{http://dx.doi.org/10.1103/PhysRevD.91.103518}{{\sf
  doi:10.1103/PhysRevD.91.103518}}].

\bibitem{Blanchet:2015bia}
L.~Blanchet and L.~Heisenberg, {\it {Dipolar Dark Matter}},
  \href{http://arxiv.org/abs/1505.05146}{{\sf arXiv:1505.05146}}.

\bibitem{Bernard:2015gwa}
L.~Bernard, L.~Blanchet, and L.~Heisenberg, {\it {Bimetric gravity and dark
  matter}},  in {\em {50th Rencontres de Moriond on Gravitation: 100 years
  after GR La Thuile, Italy, March 21-28, 2015}}, 2015.
\newblock \href{http://arxiv.org/abs/1507.02802}{{\sf arXiv:1507.02802}}.

\bibitem{Mukohyama:2014rca}
S.~Mukohyama, {\it {A new quasidilaton theory of massive gravity}},
  \href{http://arxiv.org/abs/1410.1996}{{\sf arXiv:1410.1996}}.

\bibitem{Lagos:2015sya}
M.~Lagos and J.~Noller, {\it {New massive bigravity cosmologies with double
  matter coupling}},  \href{http://arxiv.org/abs/1508.05864}{{\sf
  arXiv:1508.05864}}.

\bibitem{Matas:2015qxa}
A.~Matas, {\it {Cutoff for Extensions of Massive Gravity and Bi-Gravity}},
  \href{http://arxiv.org/abs/1506.00666}{{\sf arXiv:1506.00666}}.

\bibitem{Hinterbichler:2015yaa}
K.~Hinterbichler and R.~A. Rosen, {\it {A Note on Ghost-Free Matter Couplings
  in Massive Gravity and Multi-Gravity}},
  \href{http://arxiv.org/abs/1503.06796}{{\sf arXiv:1503.06796}}.

\bibitem{deRham:2015cha}
C.~de~Rham and A.~J. Tolley, {\it {Vielbein to the Rescue?}},
  \href{http://arxiv.org/abs/1505.01450}{{\sf arXiv:1505.01450}}.

\bibitem{DeFelice:2015hla}
A.~De~Felice and S.~Mukohyama, {\it {Minimal theory of massive gravity}},
  \href{http://arxiv.org/abs/1506.01594}{{\sf arXiv:1506.01594}}.

\bibitem{Hinterbichler:2012cn}
K.~Hinterbichler and R.~A. Rosen, {\it {Interacting Spin-2 Fields}},  {\sl
  JHEP} {\bf 1207} (2012) 047, [\href{http://arxiv.org/abs/1203.5783}{{\sf
  arXiv:1203.5783}}].

\bibitem{Gumrukcuoglu:2014xba}
A.~Emir~G{\"u}mr{\"u}k{\c c}{\"u}o{\u g}lu, L.~Heisenberg, and S.~Mukohyama,
  {\it {Cosmological perturbations in massive gravity with doubly coupled
  matter}},  {\sl J. Cosm. Astropart.} {\bf 1502} (2015) 022,
  [\href{http://arxiv.org/abs/1409.7260}{{\sf arXiv:1409.7260}}],
  [\href{http://dx.doi.org/10.1088/1475-7516/2015/02/022}{{\sf
  doi:10.1088/1475-7516/2015/02/022}}].

\bibitem{deRham:2013qqa}
C.~de~Rham, L.~Heisenberg, and R.~H. Ribeiro, {\it {Quantum Corrections in
  Massive Gravity}},  {\sl Phys.Rev.} {\bf D88} (2013) 084058,
  [\href{http://arxiv.org/abs/1307.7169}{{\sf arXiv:1307.7169}}],
  [\href{http://dx.doi.org/10.1103/PhysRevD.88.084058}{{\sf
  doi:10.1103/PhysRevD.88.084058}}].

\end{thebibliography}\endgroup

\end{document}